%% file: xe100instrument_rev.tex
\documentclass[final,3p,times,twocolumn]{elsarticle}

\usepackage{graphicx}

\usepackage{amssymb}

\usepackage{color}

\journal{Astroparticle Physics}

\begin{document}

\begin{frontmatter}



\title{The XENON100 Dark Matter Experiment}

\author[add:columbia]{E.~Aprile\corref{cor2}}
\author[add:ucla]{K.~Arisaka}
\author[add:lngs]{F.~Arneodo}
\author[add:zurich]{A.~Askin}
\author[add:zurich]{L.~Baudis}
\author[add:zurich]{A.~Behrens}
\author[add:ucla]{E.~Brown}
\author[add:coimbra]{J.M.R.~Cardoso}
\author[add:columbia]{B.~Choi}
\author[add:ucla]{D.~Cline}
\author[add:lngs]{S.~Fattori}
\author[add:zurich]{A.D.~Ferella}
\author[add:columbia]{K.L.~Giboni}
\author[add:zurich]{A.~Kish}
\author[add:ucla]{C.W.~Lam}
\author[add:columbia]{R.F.~Lang}
\author[add:columbia]{K.E.~Lim}
\author[add:coimbra]{J.A.M.~Lopes}
\author[add:zurich]{T.~Marrod\'an Undagoitia}
\author[add:rice]{Y.~Mei}
\author[add:columbia]{A.J.~Melgarejo Fernandez}
\author[add:columbia]{K.~Ni\fnref{fn1}}
\fntext[fn1]{present address: Department of Physics, Shanghai Jiao Tong University, China}
\author[add:rice]{U.~Oberlack\fnref{fn2}}
\fntext[fn2]{present address: Institut f\"ur Physik, Universit\"at Mainz, Germany}
\author[add:coimbra]{S.E.A.~Orrigo}
\author[add:ucla]{E.~Pantic}
\author[add:columbia]{G.~Plante}
\author[add:coimbra]{A.C.C~Ribeiro}
\author[add:columbia,add:zurich]{R.~Santorelli\fnref{fn4}}
\fntext[fn4]{present address: CIEMAT, Madrid, Spain}
\author[add:coimbra]{J.M.F.~dos~Santos}
\author[add:rice]{M.~Schumann\corref{cor1}\fnref{fn3}}
\cortext[cor1]{marc.schumann@physik.uzh.ch}
\cortext[cor2]{age@astro.columbia.edu (Spokesperson)}
\fntext[fn3]{present address: Physik-Institut, Universit\"at Z\"urich, Switzerland}
\author[add:rice]{P.~Shagin}
\author[add:ucla]{A.~Teymourian}
\author[add:zurich]{E.~Tziaferi}
\author[add:ucla]{H.~Wang}
\author[add:columbia]{M.~Yamashita\fnref{fn5}}
\fntext[fn5]{present address: Kamioka Observatory, ICRR, University of Tokyo, Japan}
\author{\\(XENON100 Collaboration\fnref{fn0})}
\fntext[fn0]{List includes XENON100 member institutions as of March 2009.}

\address[add:columbia]{Department of Physics, Columbia University, New York, NY 10027, USA}
\address[add:zurich]{Physik-Institut, Universit\"at Z\"urich, 8057 Z\"urich, Switzerland}
\address[add:lngs]{INFN -- Laboratori Nazionali del Gran Sasso, 67010 Assergi, Italy}
\address[add:coimbra]{Department of Physics, University of Coimbra, R.~Larga, 3004-516, Coimbra, Portugal}
\address[add:rice]{Department of Physics \& Astronomy, Rice University, Houston, TX, 77251, USA}
\address[add:ucla]{Department of Physics \& Astronomy, University of California, Los Angeles, CA, 90095, USA}

\input{abstract2}


\end{frontmatter}

\sloppy

\tableofcontents

\input{introduction2}

\input{detector2}

\input{rawdata}

\input{calibration}

\section{Outlook}

The XENON100 detector is currently operating in stable conditions underground at LNGS. After 
providing the the most stringent limits on spin-independent WIMP-nucleon scattering cross
sections for WIMP masses above $\sim$10~GeV/$c^2$~\cite{ref::xe100run08},
more data are being acquired with improved background conditions and with a lower 
trigger threshold in order to reach the full sensitivity of the instrument. 
The next generation instrument, XENON1T, with a total 
mass of $\sim$2500~kg of LXe and with 1000~kg in the fiducial target, is already in the 
technical design phase. XENON100 will be kept operational 
and running in parallel during the construction phase of XENON1T. 

An upgrade of the XENON100 detector is being considered with the goal of increasing the light collection, the operating drift field, and decreasing the overall background by further reduction of the Kr~concentration in the LXe.
The upgrade could also be useful to test new technologies required for XENON1T.

\section*{Acknowledgments}

We gratefully acknowledge support from NSF, DOE, SNF, the Volkswagen Foundation, FCT, and STCSM.
We are grateful to LNGS for hosting and supporting the XENON program, and especially to the 
LNGS mechanical workshop for their support during the construction and installation of XENON100.
We also acknowledge the support from the LNGS electronics and chemistry workshops,
computing department, and engineering team. We would like to thank the many colleagues who have 
contributed to the XENON100 construction phase, in particular Dr.~T.~Haruyama.


\input{references}
\end{document}

%% file: abstract2.tex
\begin{abstract}
The XENON100 dark matter experiment uses liquid xenon (LXe) in a time projection chamber (TPC) to search for xenon nuclear recoils resulting from the scattering of dark matter Weakly Interacting Massive Particles (WIMPs). 
In this paper we present a detailed description of the detector design and present performance results, as established during the commissioning phase and during the first science runs.

The active target of XENON100 contains 62~kg of LXe, surrounded by an LXe~veto of 99~kg, both instrumented with photomultiplier tubes (PMTs) operating inside the liquid or in xenon gas. The LXe target and veto are contained in a low-radioactivity stainless steel vessel, embedded in a passive radiation shield and
is installed underground at the Laboratori Nazionali del Gran Sasso (LNGS), Italy. The experiment has recently published results from a 100~live-days dark matter search. The ultimate design goal of XENON100 is to achieve a spin-independent WIMP-nucleon scattering cross section sensitivity of $\sigma = 2 \times 10^{-45}$~cm$^2$ for a 100~GeV/c$^2$ WIMP. 

\end{abstract}

%% file: introduction2.tex
\section{Introduction}\label{sec::introduction}

There is overwhelming observational evidence that about 23\% of the matter and energy in the universe consists
of cold dark matter \cite{ref::PDG2010,ref::bertone2005,ref::jarosik2011}, whose nature is still unknown and the subject of many investigations. Weakly Interacting Massive Particles (WIMPs) are a well-motivated class of dark matter candidates. WIMPs arise naturally in several models of physics beyond the Standard Model \cite{ref::bertone2005,ref::cheng2002,ref::bh2004}, for example in supersymmetric (SUSY) models where the lightest particles are among the most favored WIMP candidates \cite{ref::bottino2004,ref::ellis2005}. They might be observed in terrestrial experiments, sensitive enough to measure the low-energy nuclear recoil resulting from the scattering of a WIMP with a nucleus~\cite{ref::directdet}. 

The XENON dark matter project searches for nuclear recoils from WIMPs scattering off xenon nuclei. In a phased approach, experiments with increasingly larger mass and lower background are being operated underground, at the INFN Laboratori Nazionali del Gran Sasso (LNGS) in Italy~\cite{ref::LNGS}, to probe WIMP-nucleon scattering cross-sections predicted by favored SUSY models~\cite{ref::Buchmueller}. The extraordinary sensitivity of XENON to dark matter is due to the combination of a large, homogeneous volume of ultra pure liquid xenon (LXe) as WIMP target, in a detector which measures not only the energy, but also the three spatial coordinates of each event occurring within the active target. Given the rapidly falling recoil energy spectrum from WIMP interactions, and the very low interaction cross sections predicted, the challenges for XENON, as for all direct detection experiments, are to achieve a very low radioactive background and energy threshold. 

The XENON detectors are two-phase (liquid-gas) time projection chambers (TPCs), with simultaneous detection of the Xe scintillation light (S1) at the few keV$_\textnormal{\footnotesize ee}$ level (keV electron equivalent~\cite{ref::aprile2006}), and ionization (S2) at the single electron level. For a recent review of the properties of LXe as scintillator and ionizer we refer to~\cite{ref::RMP} and references therein.  The ratio S2/S1 produced by a WIMP (or neutron) interaction is different from that produced by an electromagnetic interaction, allowing a rejection of the majority of the gamma and beta particle background with an efficiency around 99.5\% at 50\% nuclear recoil acceptance. The event localization with millimeter spatial resolution and the self-shielding capability of the LXe enable further background suppression by selection of a fiducial volume. To demonstrate the XENON detector concept, the R\&D phase ~\cite{ref::aprile2006,ref::aprile2004,ref::aprile2006a,ref::ni2006,ref::aprile2005} culminated with a 10~kg scale TPC prototype (XENON10), operated at LNGS from 2006--2007~\cite{ref::xe10instrument}. XENON10 achieved some of the best limits on WIMP dark matter reported to-date~\cite{ref::xe10_si,ref::xe10_sd,ref::xe10_idm,ref::xe10_s2only}.
The ZEPLIN-II~\cite{ref::zepii} and ZEPLIN-III~\cite{ref::zepiii} experiments, conceived before XENON10, also employ the two-phase LXe TPC principle. They differ, however, in many details especially in the light detection and the background level. 

In order to increase the sensitivity to the WIMP-nucleon scattering cross section by more than one order of magnitude with respect to the state-of-the-art in 2007, a new TPC with a factor of~10 more mass and a factor of~100 less electromagnetic background was designed to fit inside the improved passive shield built at LNGS for XENON10. By focusing on  the detector's performance, the goal of a fast realization of the new and improved XENON100 experiment was successfully achieved.

Initial results~\cite{ref::xe100_11d,ref::xe100_pl} from XENON100, obtained from only 11~days of data acquired during the commissioning period at the end of 2009, have demonstrated~\cite{ref::xe100_mc} a background rate which is indeed a factor~100 less than that of XENON10. This was accomplished by careful selection of all detector materials regarding intrinsic radioactivity \cite{ref::xe100_screening}, a xenon target with lower $^{85}$Kr contamination, a novel detector design leaving only low radioactive components close to the target, and by improving the passive shield. Finally, XENON100 features an active LXe veto and allows for tighter fiducial volume cuts while still retaining a sizeable target mass. New parameter space has been excluded, competing with the limits on spin-independent WIMP-nucleon scattering cross section obtained from the full exposure of the CDMS-II experiment~\cite{ref::CDMS2}.
At the time of writing, XENON100 has set the most stringent limit for a very large range of WIMP masses ~\cite{ref::xe100run08}, and is currently the only LXe TPC
in operation with a sensitivity reach of $2 \times 10^{-45}$~cm$^2$ at 100~GeV/$c^2$ within 2012, and with a realistic WIMP discovery potential.

In this paper, we describe the design of the XENON100 detector and associated systems, and present results on its performance as established in the commissioning phase which concluded with the data reported in~\cite{ref::xe100_11d}. Following a brief summary of the operating principle of the XENON two-phase TPC, the specific design choices and  implementation in the XENON100 experiment are detailed in Section~3. Section~4 deals with raw data processing and basic-level data analysis, followed by results from calibration runs in Section~5. The paper closes with an outlook in Section~6.

\section{Principle of the XENON Two-Phase TPC}

\begin{figure}[t!]
  \includegraphics*[width=0.47\textwidth]{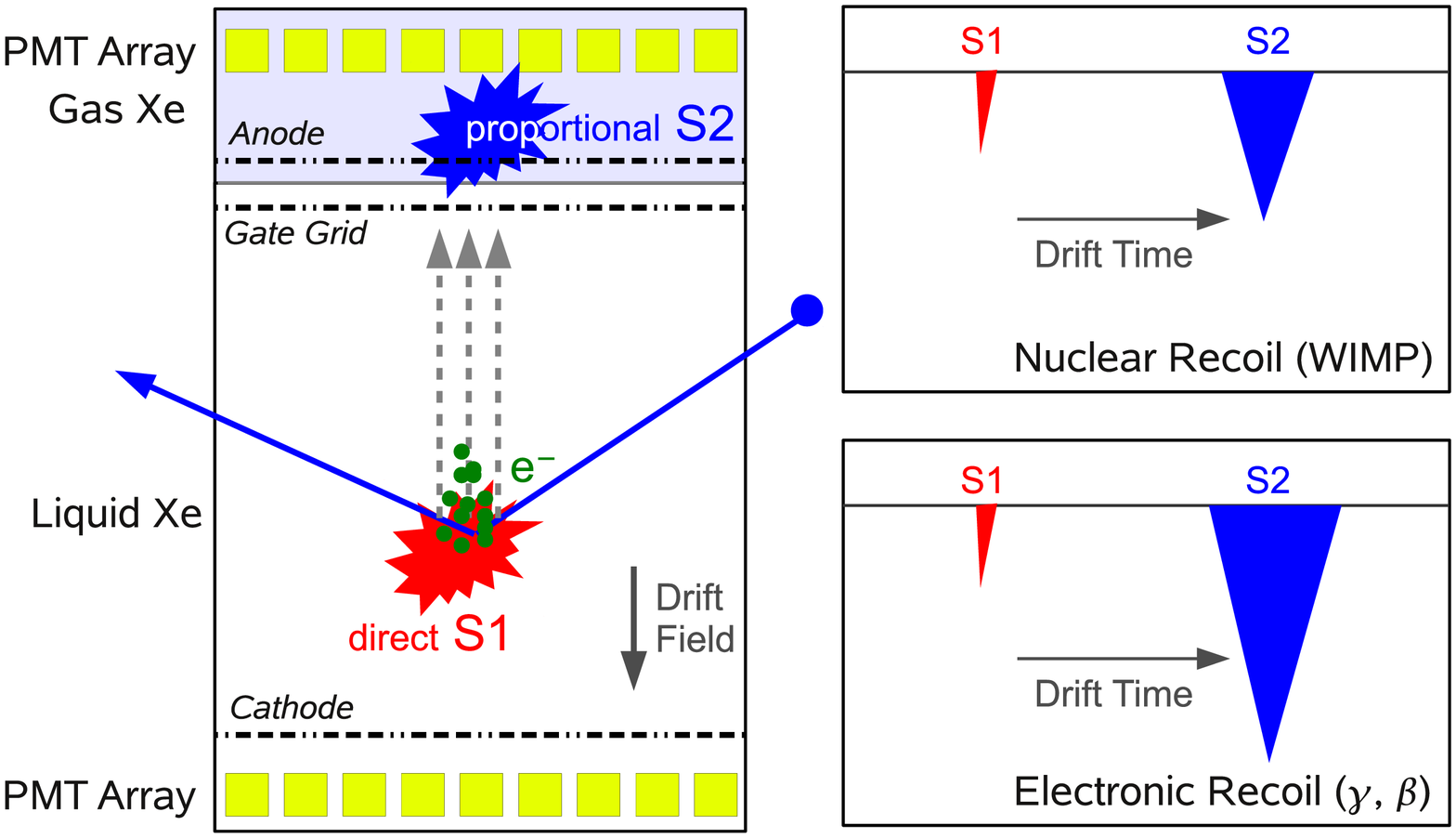}
  \caption{(Left) Working principle of the XENON two-phase liquid-gas  time projection chamber (TPC). See text for details. (Right) Sketch of the waveforms of two type of events. The different ratio of the charge (S2) and the light (S1) signal 
allows for the discrimination between nuclear recoils from WIMPs and neutrons and electronic recoils from gamma- and beta-background.}\label{fig::twophase}
\end{figure}

A schematic of the XENON two-phase (liquid-gas) time projection chamber (TPC) is shown in Fig.~\ref{fig::twophase}. A particle interaction in the liquid xenon (LXe) produces direct scintillation photons and ionization electrons. An electric field is applied across the LXe volume 
with appropriate potentials on a series of electrodes, drifting ionization electrons away 
from the interaction site. Electrons which reach the liquid-gas interface are extracted into the Xe gas, where the process of proportional scintillation takes place~\cite{ref::dolgoshein1970,ref::bolozdynya1995,ref::bolozdynya1999}. Both the direct (S1) and the proportional (S2) scintillation light, with 178~nm wavelength, are detected by photomultiplier tubes (PMTs) with optimized response in the vacuum ultraviolet (VUV) regime.

The electric field in the LXe volume is produced between a cathode at negative potential and a grounded gate grid, a few~mm below the liquid-gas interface, see Fig.~\ref{fig::twophase}. A stronger electric field in the Xe gas above the liquid is produced between the gate grid and an anode grid placed a few~mm above the liquid-gas interface. For a field larger than 10~kV/cm in the Xe gas, the electron extraction yield 
is close to 100\%~\cite{ref::aprile2004,ref::gushchin1979}.

The time difference between the S1 and the S2 signals, caused by the finite electron drift velocity in LXe at the given drift field~\cite{ref::RMP,ref::miller1968}, is proportional to the $z$-coordinate (measured
along the drift field direction) of the interaction vertex. The $x$- and $y$-coordinates can be inferred from the proportional scintillation hit pattern on the PMTs placed in the gas (top array). Thus, the XENON TPC  provides full 3-dimensional vertex reconstruction on an event-by-event basis allowing 
for the fiducialization of the target to reduce radioactive backgrounds. 

The different S2/S1 ratio of signals produced by electronic recoils (from gamma and beta background events) and by nuclear recoils (from  WIMPs and neutrons) provides additional background discrimination \cite{ref::aprile2006,ref::xe10_si}. The level of discrimination is found to be dependent on energy and electric field strength \cite{ref::aprile2005} and continues to be subject of experimental investigations.

%% file: detector2.tex
\section{The XENON100 Experiment}\label{sec::detector}

The design goal of XENON100 was to increase the target mass by a factor of ten
with respect to XENON10, and to achieve an electromagnetic background
reduction of two orders of magnitude. In this section, we give a detailed
description of the detector design and its realization, including all relevant
sub-systems.

\subsection{Detector Design}\label{sect::design}

\begin{figure}[b!]
  \includegraphics*[width=0.48\textwidth]{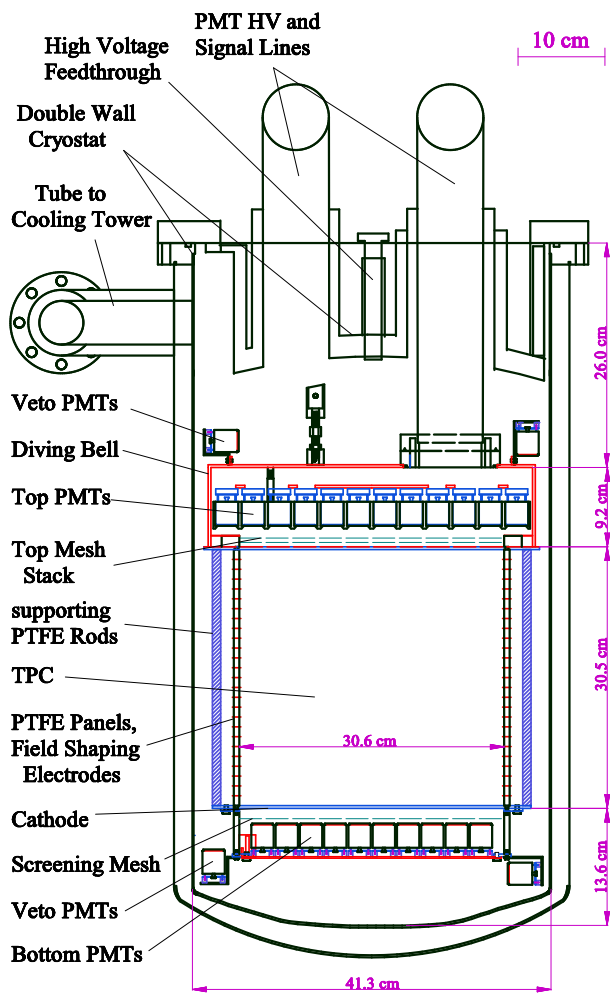}
  \caption{Drawing of the XENON100 dark matter detector: The inner TPC contains
62~kg of liquid xenon as target and is surrounded on all sides by an active liquid xenon veto of 99~kg. 
The diving bell assembly allows for keeping the liquid-gas interface at a precise level, 
while enabling to fill LXe in the vessel to a height above the bell. \label{fig::xe100cad}}
\end{figure}

The almost cylindrical XENON100 TPC of 30.5~cm height and of 15.3~cm radius contains the
62~kg LXe target (see~Fig.~\ref{fig::xe100cad}). The walls delineating the cylindrical 
volume and separating it from an active LXe veto shield, which is surrounding the 
target, are made of 24 panels of 1/4~inch-thick polytetrafluorethylen (PTFE, Teflon). 
PTFE is chosen for its properties both as insulator and good reflector for the 
VUV~scintillation light~\cite{ref::yamashita2004}. 
When cooled down to the LXe temperature of $-91^\circ$C, the PTFE panels shrink by about 1.5\%. 
To avoid scintillation light to leak 
from the active target volume to the shield region, the panels are made interlocking. 
The TPC is closed on the bottom by the cathode, and on the top by the gate grid (see~Sect.~\ref{sec::efields}). 

The two-phase (liquid-gas) operation requires a precisely controlled liquid level just covering the gate grid. To minimize the impact of liquid density variations due to temperature changes as well as 
fluctuations in the gas recirculation rate, a diving bell design was chosen to keep the liquid at a precise level. Outside the bell, the liquid in the detector vessel can be at an arbitrarily high level. This made it possible to fill the vessel to a height of about 4~cm above the bell, enabling a $4\pi$ coverage of the TPC with a LXe veto.  

The bell keeps the liquid level at the desired height when a constant stream of gas pressurizes it. This is accomplished by feeding the xenon gas returning from the gas recirculation system (see~Sect.~\ref{sec::lxepurification}) into the bell. The pressure is released through a small pipe that reaches out into the veto LXe volume. The height of the LXe level inside the bell is adjusted by vertically moving the open end of the pipe which is connected to a motion feedthrough.

In order to minimize the dependence of the charge signal on the $xy$-position, 
the liquid-gas interface has to be parallel to the anode. 
To facilitate leveling, the detector can be tilted with two set screws from 
the outside of the radiation shield.
Four level meters, measuring the capacitance between partially LXe filled stainless
steel tubes and a Cu rod placed in their center, as well as the measured S2~signal width at different 
locations, are used to level the detector (see Sect.~\ref{sec::level}).

\begin{figure}[b!]
  \centering
  \includegraphics*[width=0.4\textwidth]{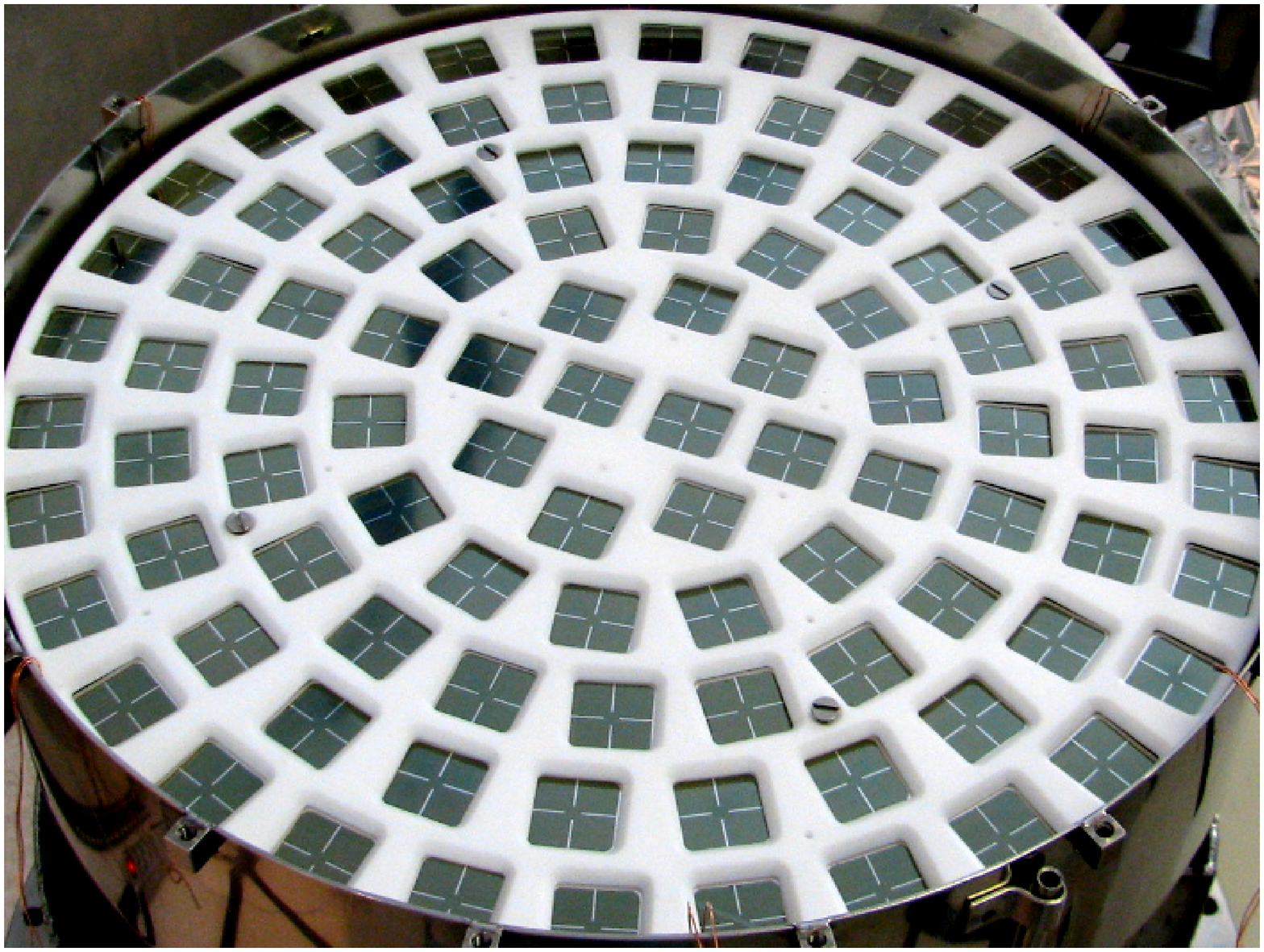}
  \includegraphics*[width=0.4\textwidth]{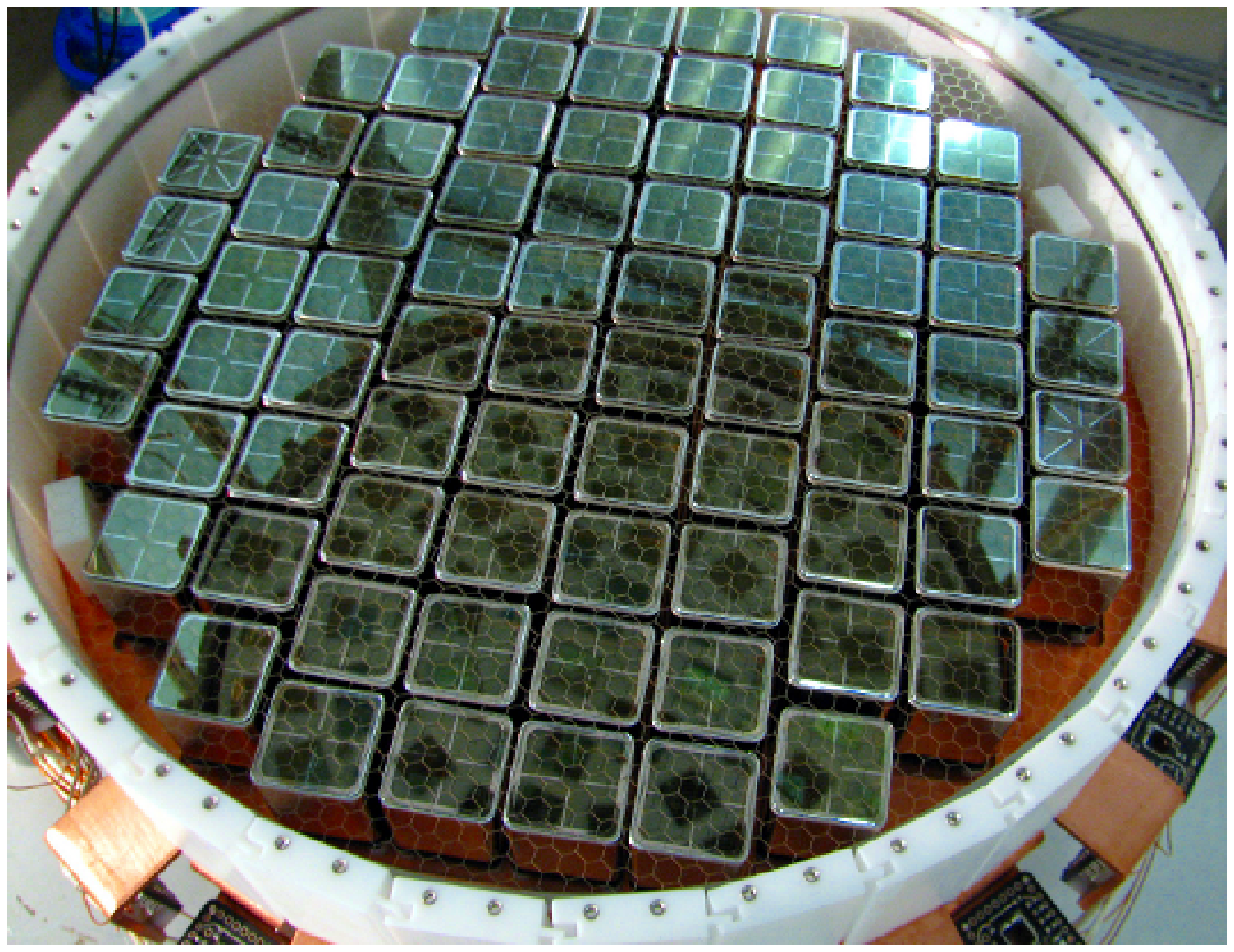}
  \caption{(Top) The Hamamatsu R8520-06-Al PMTs on the top of XENON100 are arranged in
concentric circles in order to improve the reconstruction of the radial event position. (Bottom)
On the bottom, the PMTs are arranged as closely 
as possible in order to achieve high light collection, as required for a low 
detector threshold.}
\label{fig::pmts}
\end{figure}

\begin{figure}[t]
  \centering
  \includegraphics*[width=0.48\textwidth]{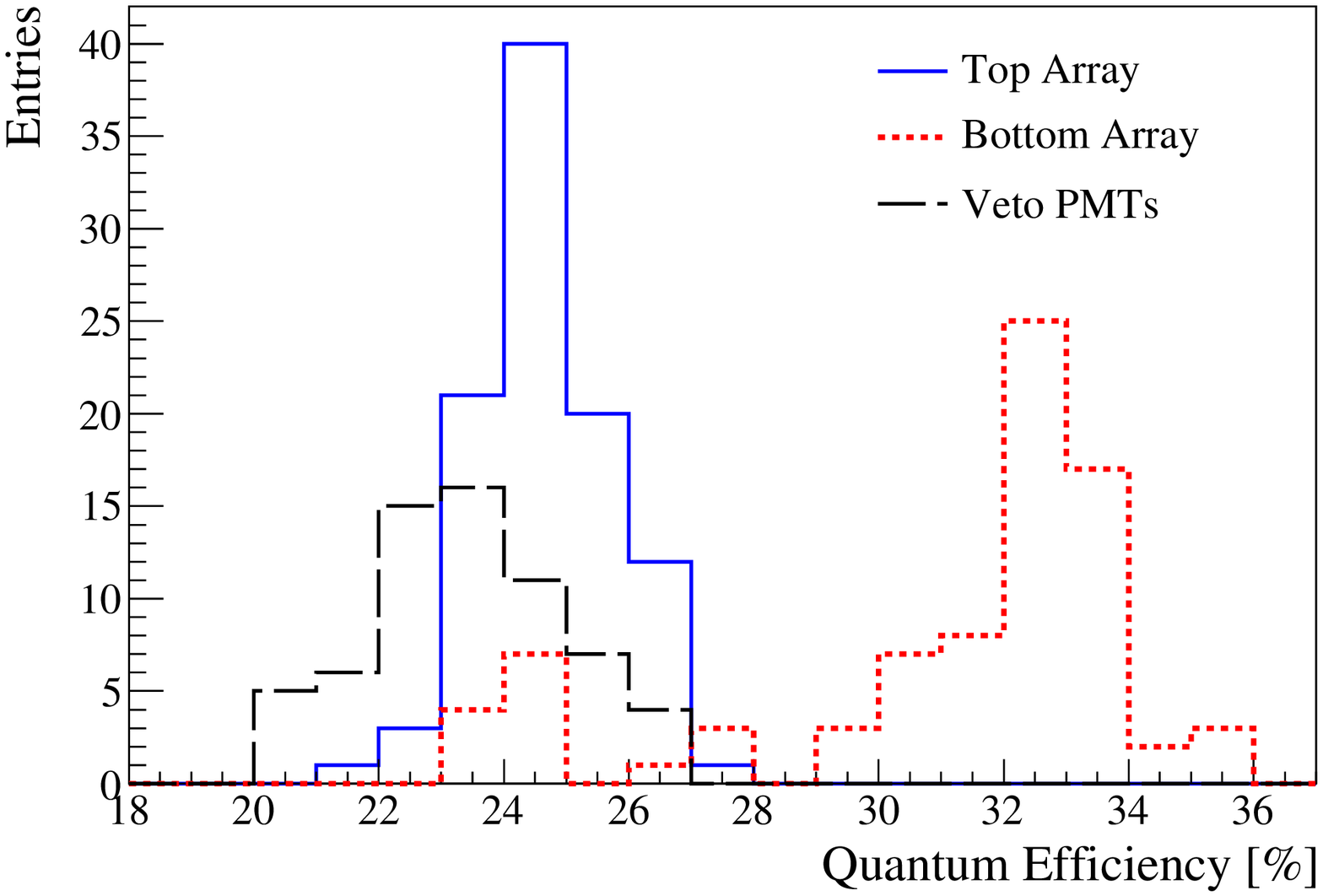} \vspace{2mm}
  \includegraphics*[width=0.48\textwidth]{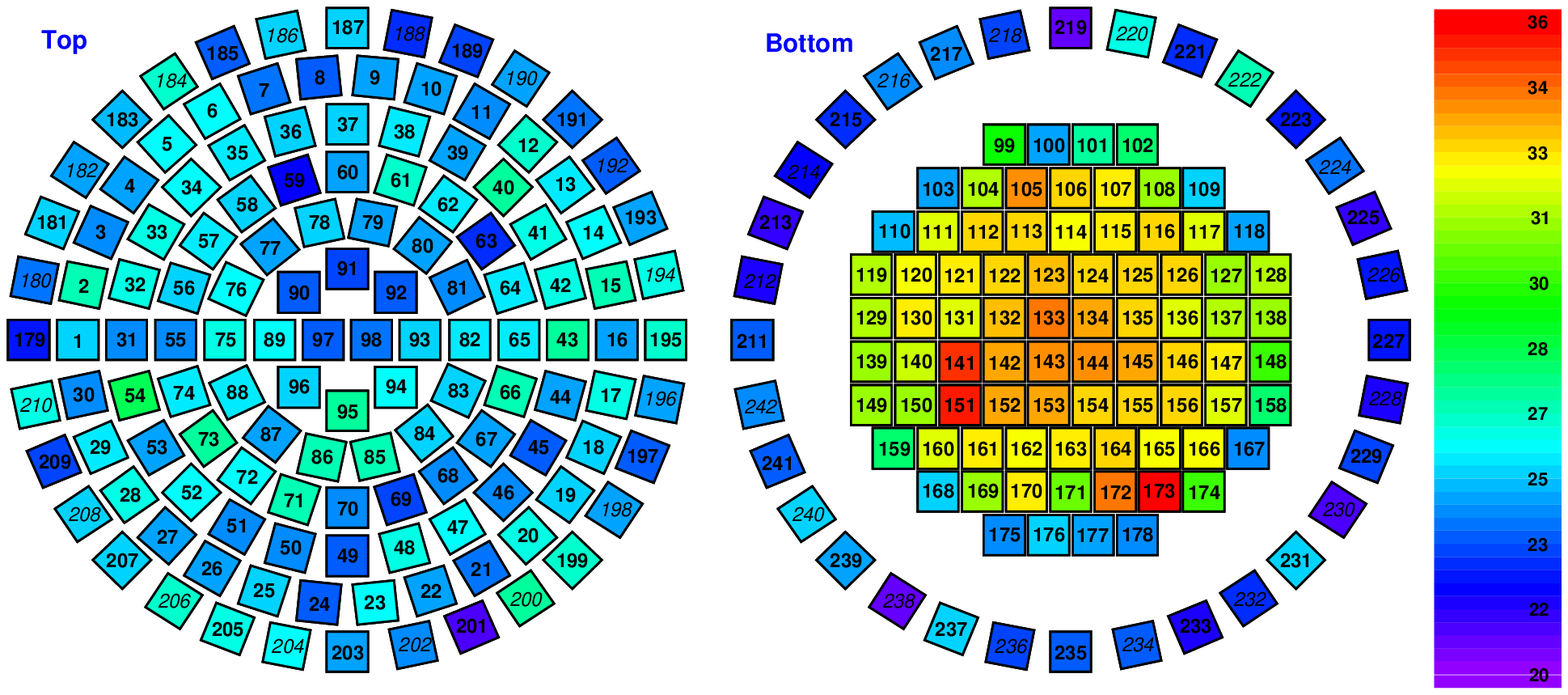}
  \caption{(Top) Quantum efficiency (QE) distribution of the R8520-06-Al PMTs in XENON100, estimated at room temperature by Hamamatsu. The PMTs on the bottom have higher values to improve the energy threshold. (Bottom) Arrangement of the PMT with different QE (given by the color code) in the detector. Tubes with lower QE are placed mainly on the top, in the active veto or in detector corners, where they have less impact on the detector threshold.}
\label{fig::qe}
\end{figure}

Two arrays of Hamamatsu R8520-06-Al 1"~square photomultiplier tubes (PMTs),
specially selected for low radioactivity~\cite{ref::xe100_screening}, detect the light in the TPC: 98~PMTs
are located above the target in the gas phase, arranged in concentric circles in order to 
improve the resolution of radial event position reconstruction, see~Fig.~\ref{fig::pmts} (top). The outmost ring 
extends beyond the TPC radius to improve position reconstruction at the edges. The remaining photocathode coverage is 43.9\% of the TPC cross section area.
The energy threshold and hence the sensitivity of the detector is determined by the S1~signal. Because 
of the large refractive index of LXe of $(1.69\pm 0.02)$~\cite{ref::solovov2004}, and the consequent
total internal reflection at the 
liquid-gas interface, about 80\% of the S1~signal is seen by the second PMT array, which is 
located below the cathode, immersed in the LXe.
Here, 80~PMTs provide optimal area coverage (in average 52\% useful PMT photocathode coverage with 61\% in the central part) for efficient S1~light collection, see~Fig.~\ref{fig::pmts} (bottom).
The bottom PMTs have a higher quantum efficiency compared to the top PMTs. This is shown in Fig.~\ref{fig::qe}, together with the distribution of the PMT quantum efficiency in the detector. The photoelectron collection efficiency from the photocathode to the first dynode for this type of PMT is about 70\%, according to Hamamatsu.

A LXe layer of about 4~cm thickness surrounds the TPC on all sides and 
is observed by 64~PMTs, of the same type as used for the TPC readout. In total, this volume 
contains 99~kg of LXe. The presence of this LXe veto, operated in anti-coincidence mode, is very effective for background reduction~\cite{ref::xe100_mc}
and is one major difference in design compared to XENON10. 
The LXe veto is optically separated from the TPC by the interlocking PTFE panels. 
Optical separation of target and veto has some advantages over instrumenting the entire volume as TPC as it 
lowers the event rate in the target and reduces the rate of accidental coincidences to a neglegible level.
It also requires fewer PMTs and reduces cost.

The voltage divider, directly coupled to the PMT, is a printed circuit with surface mount components on a Cirlex substrate~\cite{ref::cirlex}. Compared to XENON10~\cite{ref::xe10instrument}, the design was modified and the number of electronic components was  reduced in order to decrease the background contribution~\cite{ref::xe100_screening}.
Each PMT is connected with a PTFE insulated coaxial cable to read the signal as well as a Kapton insulated single wire cable to supply the operating voltage.
The outer insulation of the coaxial cable was removed to prevent potential impact of trapped air on LXe purity. This choice resulted in a common ground  for all PMTs.
The cables of the top PMTs are guided out of the passive shield through a pipe connecting to the bell. All other cables are grouped into 4~bunches which are fixed to the PTFE support structure, and guided on top of the bell where they are collected and routed out of the shield through a second pipe (see Fig.~\ref{fig::xe100cad}). Each pipe is equipped with commercial 48-pin vacuum feedthroughs at the end providing the connection to atmosphere, outside of the shielding.

Nine out of the 242~PMTs installed in XENON100 were not working during the physics run reported in \cite{ref::xe100run08}. Of these, four are on the top array (PMT 9, 12, 39, and 58), two on the bottom (PMT 100, 105) and three in the active veto (PMT 195, 224, 235). Three additional PMTs (148, 177, 191) were powered off due to high dark current. Overall, only 5\% of the PMTs are non-functional at the time of writing, almost 3~years after the last detector opening for maintenance.

The TPC is mounted in a double-walled
316Ti stainless steel cryostat, selected for its low activity, especially in
$^{60}$Co~\cite{ref::xe100_screening}. Since the radioactive contamination of the cryogenics system,
ceramic feedthroughs, etc.~cannot be lowered easily, the detector is cooled remotely and all parts 
with a known high radioactive contamination are installed away from the detector itself, outside 
the passive shield (see Sect.~\ref{sect::detshield} and Fig.~\ref{fig::xe100shield}).
The connection to the outside of the shield is established via
3~stainless steel pipes, one double-walled to the cooling system, the
others single-walled to the PMT feedthroughs and pumping ports. 
To bias the cathode and the anode, custom-made hermetic HV feedthroughs, of similar design 
as developed for XENON10~\cite{ref::xe10instrument}, are used also for XENON100. They are 
made of a stainless steel core with a PTFE insulation layer, for reduced radioactive contamination 
compared to commercial HV feedthroughs. 

PTFE has a rather large linear thermal expansion coefficient $A \! \sim \! 1.2 \! \times \! 10^{-4}$~K$^{-1}$, as
measured for the PTFE used in XENON100. This leads to a TPC contraction of $\sim$5~mm when cooled to LXe temperatures. The contraction along the $z$-dimension is taken into
account when the $z$-coordinate of an event is determined. 
Radial contraction is negligible since the PTFE panels are mounted between copper support rings which have a 
much smaller thermal expansion coefficient ($A_{Cu}\! \sim \! 1.5 \! \times \! 10^{-5}$~K$^{-1}$).

\subsection{Electric Field Configuration}\label{sec::efields}

Thin metal meshes are used to create the electric fields required to operate XENON100
as a two-phase TPC. They were chemically etched from stainless steel foils and spot-welded 
onto rings made of the same low radioactivity stainless steel that is used for the cryostat. Before 
welding, the meshes were stretched in order to minimize sagging.

The cathode mesh is 75~$\mu$m thick with a hexagonal pattern and a pitch of 5~mm. 
A grounded screening mesh, also of hexagonal pattern and 5~mm pitch, but
50~$\mu$m thick, is placed 12~mm below the cathode,
and 5~mm above the bottom PMTs to shield them from the cathode high voltage. 
According to the initial design,
a voltage of $-$30~kV would bias the cathode, to generate a drift field of 1~kV/cm corresponding to a maximal
electron drift time of $\sim$160~$\mu$s. However, due to 
the appearance of small light pulses at increasingly high cathode voltage, the voltage was reduced to $-$16~kV for stable operation, resulting in a drift field of 0.53~kV/cm across the TPC. The pulses are most likely caused by electron field emission
and subsequent scintillation in the strong electric field near sharp features of the cathode mesh.

The unavoidable liquid layer between the cathode and the bottom PMTs is a charge insensitive 
region and a potential source of events which can be confused with true nuclear recoils. For example,
 a background gamma-ray with two interactions, one in this insensitive region and one inside the TPC 
active region, may result in a reduced S2/S1~ratio, mimicking a nuclear recoil event (see e.g.~\cite{ref::xe10instrument,ref::xe10_si}). These events mainly occur close to the cathode and can be 
reduced by a $z$-cut. In addition, the distribution of the S1~light on the PMTs differs from real 
single scatter interactions and this is exploited in data analysis~\cite{ref::xe10instrument,ref::xe100run08}.  

About 15~mm below the top PMTs, the TPC is closed with a
stack of 3~stainless steel meshes with hexagonal pattern: 
a central anode (125~$\mu$m thick, 2.5~mm pitch)
between two grounded meshes with a spacing of 5~mm. An extraction field of
$\sim$12~kV/cm is obtained by applying $+$4.5~kV to the anode. The exact
value of the extraction field depends on the position of the liquid-gas
interface because of the different dielectric constants of LXe and Xe gas. The field is high enough to obtain an extraction efficiency close to 100\%~\cite{ref::aprile2004,ref::gushchin1979}. 
The grounded mesh above the anode shields the amplification region from external fields and
yields more homogeneous proportional scintillation signals.

In order to optimize the S2 performance, the anode could be moved horizontally with respect
to the gate grid and the top grid. It was aligned at a half-pitch offset under a microscope and
fixed with set screws. The whole stack is optimized for optical transparency and
minimal impact on the S2 energy resolution.
The spread of the S2 signal due to the varying electron path length is only~4\%, 
independent of the S2 energy. 
Averaged over all angles of incidence, the optical transparency of the top mesh
stack and of cathode plus screening mesh is 47.7\% and 83.4\%, respectively.

A homogeneous electric field across the $\sim$30~cm long TPC drift gap is created by a field 
cage structure made of thin copper wires. Two wires, at the same potential, one 
running on the inside and one on the outside of the PTFE panels, are used to 
emulate a 1/4~inch-wide field shaping electrode which generates the desired straight field lines 
within the target volume. 40~equidistant field shaping electrodes, connected through 700~M$\Omega$ 
resistors are used. 

The penetration of electric field lines through the cathode, facilitated by the large mesh pitch 
and the thin wire diameter chosen to optimize light collection, distorts the electric field 
at large radii, just above the cathode. The correction of this effect, based on calibration data, 
is presented in Sect.~\ref{sec::efieldCorr}.

\subsection{The Passive Shield}\label{sect::detshield}

The XENON100 experiment is installed underground at LNGS, at the same site as XENON10,
in the interferometer tunnel away from 
the main experimental halls. At the depth of 3700~m water equivalent, the surface muon 
flux is reduced by a factor $10^6$~\cite{ref::lngsmuon}. 

\begin{figure}[h!]
  \centering
  \includegraphics*[width=0.48\textwidth]{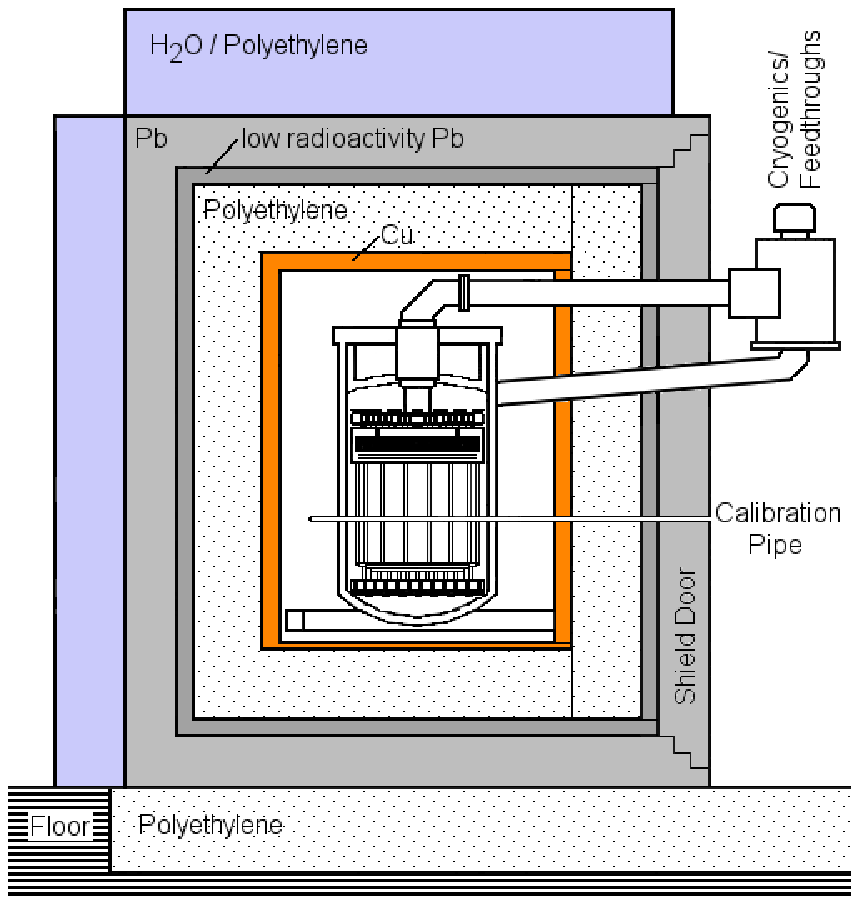}\vspace{1mm}
  \includegraphics*[width=0.40\textwidth]{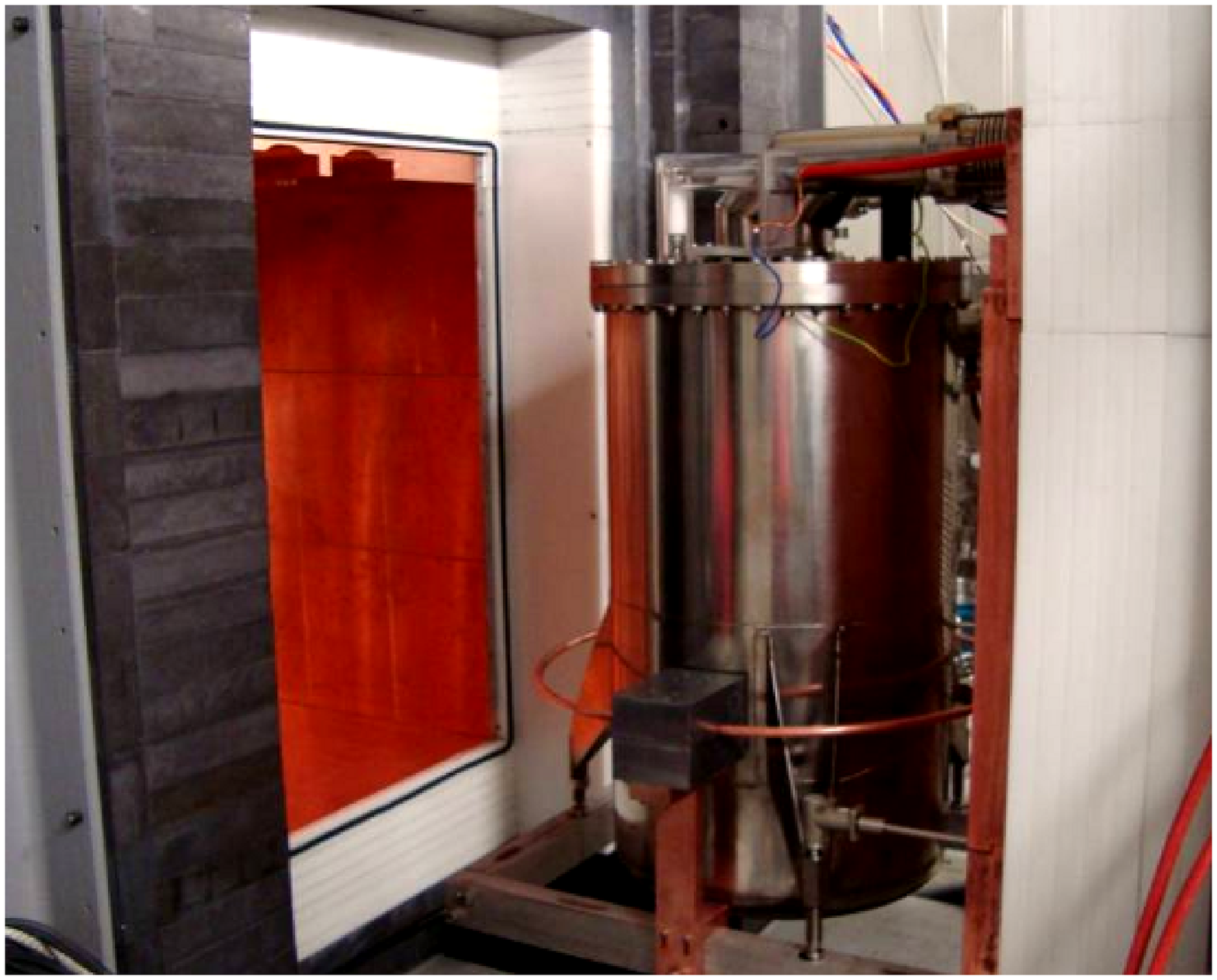}
  \caption{(Top) Drawing of the XENON100 detector in its passive shield made out of
copper, polyethylene, lead, and water containers. (Bottom) XENON100 with
opened shield door. The Pb-brick along the calibration pipe around the cryostat is a gamma-shield during $^{241}$AmBe neutron calibrations.}\label{fig::xe100shield}
\end{figure}

In order to reduce the background from the radioactivity in the experiment's environment, in the laboratory walls etc.~\cite{ref::haffke}, additional passive shielding is needed. An improved
version of the XENON10 shield \cite{ref::xe10instrument} was required in light of 
the increased sensitivity of the XENON100 experiment. The detector is surrounded (from inside to outside) by 
5~cm of OFHC copper, followed by 20~cm of polyethylene, and
20~cm of lead, where the innermost 5~cm consist of lead with a low $^{210}$Pb contamination of $(26\pm6)$~Bq/kg~\cite{ref::xe100_screening}.
The entire shield rests on a 25~cm thick slab of polyethylene.
An additional outer layer of 20~cm of water or polyethylene has been added on 
top and on 3~sides of the shield to reduce the neutron background further. 
Fig.~\ref{fig::xe100shield} shows a sketch of XENON100 inside the shield.

During detector operation, the inner shield cavity is constantly purged with
high purity boil-off nitrogen at a rate of 17~standard liters per minute (SLPM) in 
order to avoid radioactive radon penetrating into the shield. The remaining radon 
concentration is constantly monitored with a commercial radon detector 
and is at the limit of the detector's sensitivity \mbox{($<1$~Bq/m$^3$)}.

Careful selection of materials to build the detector is crucial 
in order to reach a competitive dark matter detection sensitivity. All components
used for the XENON100 detector and shield have been chosen according
to their measured low intrinsic radioactivity. These measurements were performed
using a dedicated screening facility~\cite{ref::gator}, consisting of 
a 2.2~kg high purity Ge detector in an ultra-low background Cu cryostat and
Cu/Pb shield, operated at LNGS, as well as the LNGS screening 
facilities~\cite{ref::lngsscreening1,ref::lngsscreening2}. 

The electronic recoil background of XENON100 is dominated by gamma rays from the decay chains
of radioactive contaminants, mostly $^{238}$U, $^{232}$Th, $^{40}$K, and
$^{60}$Co, in the detector materials. The screening results \cite{ref::xe100_screening} 
are used to predict the overall background rate. Rate and spectral shape agree very well with the measurement~\cite{ref::xe100_mc}.

\subsection{Cryogenic System}\label{sec::cooling}

\begin{figure}[t!]
  \centering
  \includegraphics*[width=0.47\textwidth]{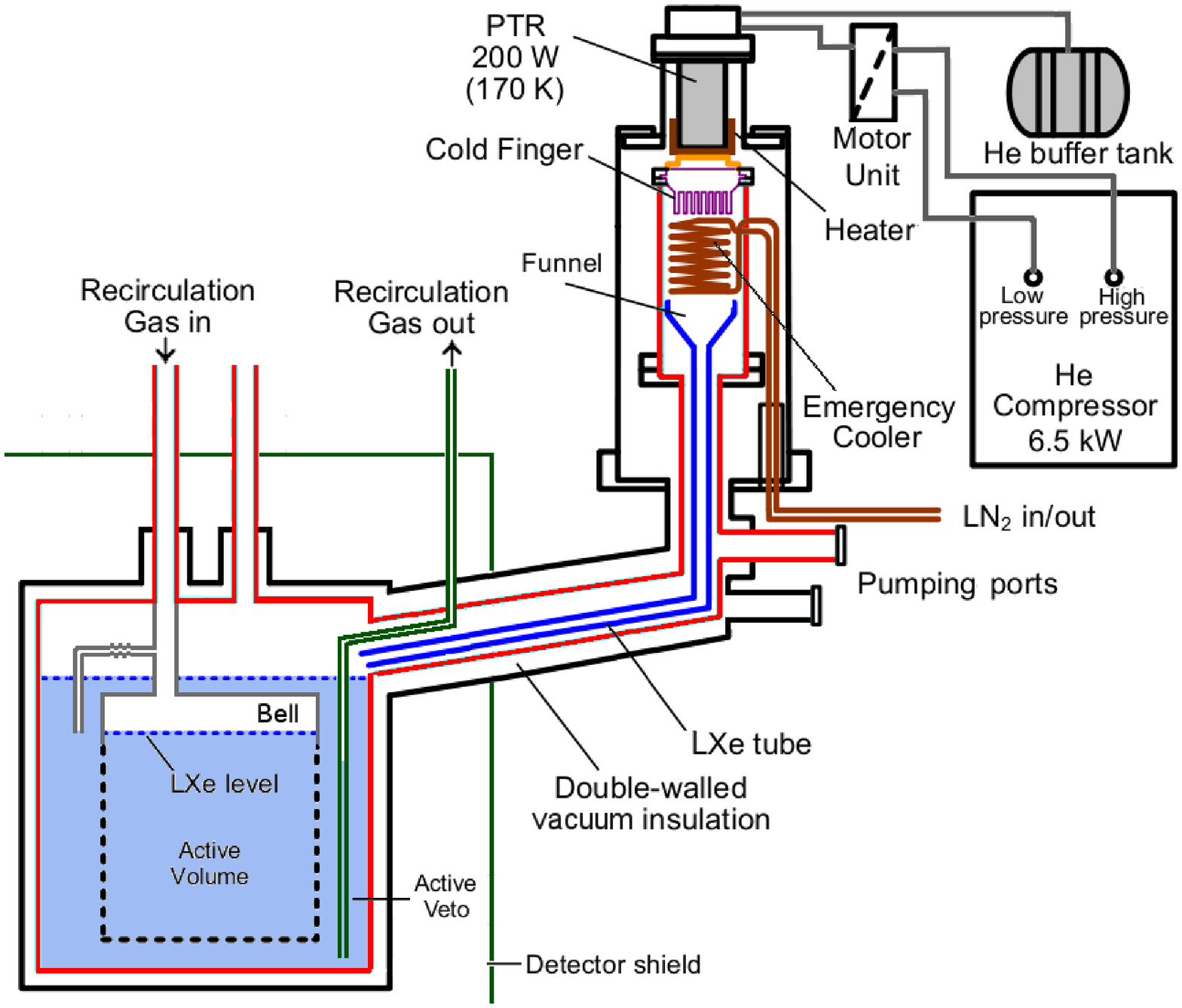}
  \caption{Sketch of the XENON100 cryogenic system. The cooling is provided by a 200~W pulse 
tube refrigerator (PTR) installed outside the shield and connected to the main cryostat via 
a double-walled vacuum insulated pipe. Xenon purification is independent of the cooling. LXe is
extracted from the bottom of the detector, purified in gas phase, and introduced back as xenon gas into the diving bell. In combination with a height-adjustable gas outlet pipe, the bell
allows for a precise control of the liquid level, while having LXe all around the detector. 
The figure is not drawn to scale and details of the TPC are omitted for clarity. }\label{fig::cryosystem}
\end{figure}

A reliable, easy to use cooling system with very good stability is needed 
for any dark matter experiment operated at cryogenic temperatures. Pulse Tube 
Refrigerators (PTRs)~\cite{ref::haruyama2006}, specifically designed for high cooling power at LXe
temperatures, were employed from the start of the XENON project. 
The PTR for XENON100 is an Iwatani PC150, driven by a 6.5~kW
helium compressor. The cooling power for this combination is measured to be 200~W at
170~K.  A schematic of the cooling system is shown in Fig.~\ref{fig::cryosystem}.

The PTR cold-head is mounted on a cylindrical copper block that closes off the inner 
detector vessel and that acts as a cold-finger. The  cold-finger is sealed to the 
inner detector vessel with a seal made of a pure aluminum wire. The PTR can thus be 
serviced or replaced without exposing the detector volume to air. A copper
cup with electrical heaters is inserted between the PTR cold-head and the 
cold-finger. The temperatures above and below the heater are measured with 
precise temperature sensors. A proportional-integral-derivative (PID)
controller regulates the heating power required to keep the temperature of the 
cold-finger, and hence the Xe vapor pressure in the detector, at the desired value. 

The PTR is mounted in a separate double-walled vacuum insulated vessel, 
placed outside the passive shield, along with many auxiliary modules, including the 
motor valve and buffer tank, which have to be within 50~cm of the PTR cold-finger for optimal performance (see~Fig.~\ref{fig::cryosystem}). The bottom of this ``cooling tower'' is 
connected to the main cryostat with a vacuum insulated pipe at a height above the 
liquid level. The boil-off Xe gas from the detector can thus reach the cold-finger 
of the PTR where it is liquefied. The liquid drops are collected by a funnel  
and flow back into the detector through a smaller diameter pipe 
at the center of the insulated pipe. To drive the liquid flow, the insulated pipe is 
inclined by 5$^\circ$ with respect to the horizontal. This cryogenic system design 
with the PTR assembly placed far from the detector, enabled a reduction of the background 
from radioactive materials~\cite{ref::xe100_mc}. The total mass of steel within the 
shield cavity was reduced from 180~kg for the much smaller XENON10 detector to about 
70~kg for XENON100.

In case of emergency, e.g., a prolonged power failure or a failure of the primary 
cooling system, the detector can be cooled by liquid nitrogen (LN$_2$). A stainless steel coil 
is wound around the cold finger and is connected to an external LN$_2$
dewar, always kept
full during detector operation. The LN$_2$ flow through the coil is controlled by an 
actuated valve and triggered when the detector pressure increases above a defined 
set-point. Tests have shown that the detector can be kept stable for more than 24~hours
without any human intervention using the emergency LN$_2$ cooling system.

\subsection{The Gas Handling and Purification System}\label{sec::lxepurification}

A total amount of 161~kg of LXe is necessary to fill the target volume and the active veto. 
It is stored in 4~large (75~$\ell$ volume) high-pressure aluminum gas 
cylinders, which are surrounded by custom-made insulated LN$_2$ dewars. This allows them  
to be cooled down to
recover the xenon gas from the detector by freezing the Xe in the cylinders. Both Xe filling and recovery takes 
place in gas phase, through a stainless steel pipe connecting the storage 
with the purification system (see below).  
All pipes, flow controllers, regulators, and valves are metal sealed. 
During the process of filling, the gas is liquefied by the PTR, which has sufficient cooling power 
to cool the vessel, condense the gas, and keep it at the operating temperature of $-$91$^\circ$C.
The xenon gas is liquefied at a rate of almost 3~kg/hour, limited by the cooling capacity of the PTR. 

During gas recovery, the Xe is extracted from the detector by a double-diaphragm pump and transferred to 
the storage cylinders at LN$_2$ temperature. Recovery is facilitated by breaking the vacuum insulation of the cryostat. 

During the Xe purification from Kr, through a dedicated cryogenic distillation column (see 
Sect.~\ref{sec::kr}), the gas stored in the cylinders is passed through the distillation column 
before being filled directly into the detector. To replenish the Kr-rich Xe, which is produced as 
``off-gas'' during distillation, more Xe than needed for a complete fill of XENON100 is stored 
in the cylinders.

\begin{figure}[b!]
  \centering
  \includegraphics*[width=0.4\textwidth]{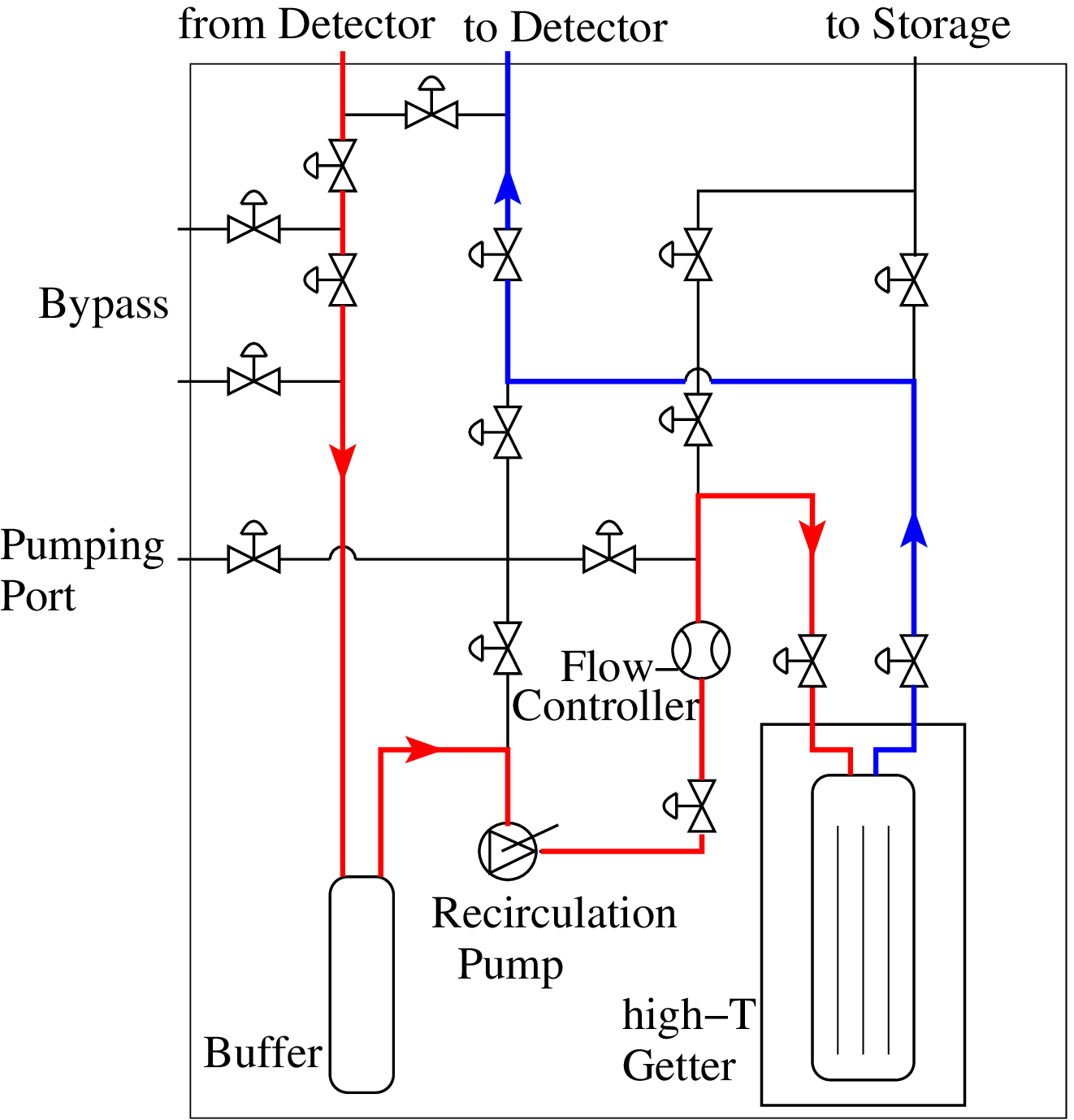}
  \caption{Schematic of the XENON100 purification system. LXe is extracted from the detector using
a diaphragm gas pump. It evaporates in the gas lines and is passed through a high temperature getter
for gas purification, before it is pushed back into the detector (the path for standard operation is indicated by the arrows). Different valves allow for the bypassing of components for special operations like detector filling or recuperation, or maintenance.
Valves to atmosphere are used to add auxiliary equipment for gas analysis or detector calibration.}
\label{fig::purification}
\end{figure}

The light and the charge signal from particle interactions in the xenon are adversely
affected by impurities in LXe: The light is mostly attenuated by
water~\cite{ref::meg_abs}, whereas for the charge the most abundant and harmful impurity is
oxygen which leads to charge losses while the electron cloud drifts towards 
the liquid-gas interface~\cite{ref::chargeabs}. Hence, oxygen and other electro-negative impurities in commercial xenon have to be removed well below the 1~ppb (part per billion) level oxygen-equivalent to achieve the
required low charge attenuation (high electron lifetime) and long VUV photon absorption length~\cite{ref::RMP}. 
The xenon is purified by constantly recirculating xenon gas through a high temperature zirconium 
getter ({\it SAES MonoTorr PS3-MT3-R/N-1/2}, see~Fig.~\ref{fig::purification}), which removes impurities by chemically bonding them to the getter material. At a rate of about 5~SLPM, liquid from
the bottom of the detector vessel is evaporated and pushed through the getter by a
double diaphragm pump ({\it KNF N143.12E}), before it is returned to the detector.

In order to speed up the purification process already before filling with LXe, the
leak-tested detector was heated to 50$^\circ$C (the temperature limit is set by
the PMTs) while the detector vacuum was monitored with a residual gas analyzer (RGA).
Since Xe is known to act as an effective solvent due to its polarizability~\cite{ref::rentzepis}, 
the detector was then
filled with 2~atm of Xe gas. The detector was heated again while the warm gas
was passed through the getter for purification for several weeks. During this
process, the decrease of the water content from $\sim$500~ppb to the 1~ppb level was monitored with
a dedicated apparatus ({\it Tigeroptics HALO}) using a spectral absorption technique.

\subsection{Light and Charge Yield Evolution}\label{sec::yieldevolution}

\begin{figure}[t!]
  \includegraphics*[width=0.47\textwidth]{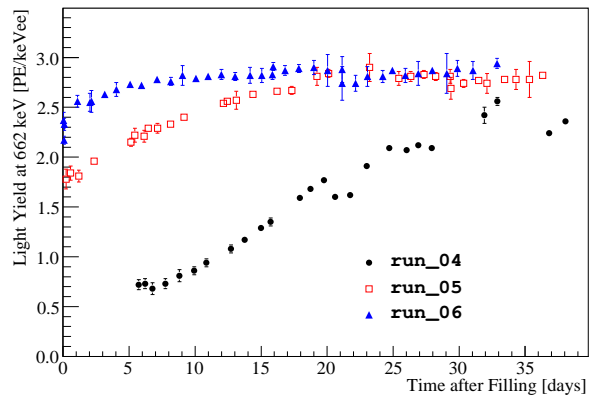}
  \caption{Improvement of the light yield vs.~time with continuous circulation through the getter,   
   for several early commissioning runs. The yield, corresponding to the value of the 662~keV 
full-absorption peak, was measured with an external $^{137}$Cs source at zero drift field and no position
dependent corrections have been applied to the data.
The maximum was reached in all cases but on different timescales due to
different and improved preparation of the detector.}\label{fig::lyevolution}
\end{figure}

Analyses of the light yield and the $z$-dependent charge yield from standard gamma
calibration sources such as the 662~keV line from $^{137}$Cs give access to the 
water and oxygen content in the LXe, respectively. 
During the initial 
commissioning phase of XENON100, the light yield of the LXe filled detector increased while 
water contamination, 
from materials outgassing, was reduced by continuous circulation of the gas through the high 
temperature getter (see Fig.~\ref{fig::lyevolution}). 
The time to reach the maximum light yield was considerably decreased between the runs. The 
detector was always exposed to atmosphere in between these runs, however, the detector preparation prior 
to filling the gas was improved, mostly by recirculation of Xe gas at a temperature of 50$^\circ$C 
for 2-3 weeks as described in Sect.~\ref{sec::lxepurification}, in order  
to clean the inner detector surfaces efficiently. 

The pipes guiding all signal and HV cables out of the passive
shield were designed to be single-walled to reduce the mass of possibly
radioactive materials close to the detector. Consequently,
their inner surfaces have temperatures of 5$^\circ$C - 20$^\circ$C during normal
operation and thus exhibit a much higher outgassing rate than the surfaces at LXe temperature. 
This effect is additionally boosted by the large
amount of wires (PTFE insulated coaxial wires for the signal and Kapton wires
for the high voltage) in the pipes, constituting a large surface. Bake-out at 
the maximum allowed temperature ($\sim$120$^\circ$C) and the continuous 
circulation of Xe gas reduced the outgassing with time, leading to a constant light yield
during the XENON100 commissioning run~\cite{ref::xe100_11d} and the science run~\cite{ref::xe100run08}. 

The charge yield in LXe, at a given drift field, is strongly affected by 
electronegative impurities dissolved in the liquid. External leaks, surface outgassing,
and contaminations present in the liquid itself all may contribute. The remaining amount
of electronegative impurities must be kept at the sub-ppb level of O$_2$ 
equivalent in order to drift freely ionization electrons across large distances. With 
about 30~cm, the drift 
gap in the XENON100 TPC is twice as long as that of XENON10 and the longest of any 
ionization detector operating with LXe to-date. The stringent constraints on materials choice 
driven by radio-purity, the limited allowed temperature range of the PMTs and
their voltage-divider network, the large amount of cabling, the 
limited speed in gas purification rate determined by the total available cooling power, all 
contributed to the challenge of achieving long electron drift in this novel detector. 
Compared to the time required to reach the maximal scintillation light yield, it took much longer to 
achieve a LXe purity which allowed for electrons to drift over the 
full 30~cm gap. This was largely 
due to the dominating role of materials outgassing and due to impurities introduced by leaks 
of the vessel which developed during transport and installation underground. 

\begin{figure}[t!]
\includegraphics*[width=0.47\textwidth]{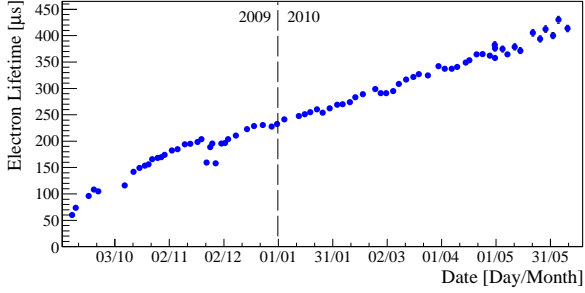}
  \caption{Evolution of the electron lifetime $\tau_e$, measured with a $^{137}$Cs source, 
during the commissioning run~\cite{ref::xe100_11d} and the first science run~\cite{ref::xe100run08}
which started on January 13, 2010. The
decrease around the end of November 2009 is due to a stop of
xenon purification during an exchange of the recirculation pump. In the meantine, values of $\tau_e>0.5$~ms have
been achieved.
}\label{fig::chargeevolution}
\end{figure}

The electron lifetime $\tau_e=(k_n n)^{-1}$, where $k_n$ is the attachment rate coefficient, 
is a measure of the number of electrons lost during 
the drift time, and thus a measure of the total impurity concentration $n$ in the liquid. 
In the XENON100 TPC, this number is regularly monitored by measuring the S2 signal of the
full-absorption peak of 662~keV gamma-rays in the fiducial volume as a function 
of drift time, using an external $^{137}$Cs source. 
The evolution of $\tau_e$ with purification time, shown in
Fig.~\ref{fig::chargeevolution} for the published science runs~\cite{ref::xe100_11d,ref::xe100run08}, 
indicates that the maximum has yet to be reached,
leading to a time-dependent correction to the data. During dark
matter data taking, this charge calibration is done regularly in order
to correct the S2~signal for charge loss, using linear interpolation between
calibrations or a linear fit (see Sect.~\ref{sec::s2corr}). At the time of writing, values larger than 
550~$\mu$s have been achieved (see Fig.~\ref{fig::elife} on page~\pageref{fig::elife}) and the maximum has not been reached yet.

\subsection{The Krypton Distillation System}\label{sec::kr}

Xenon has no long-lived radioactive isotopes and the half-life of the potential double-beta 
emitter $^{136}$Xe is so long~\cite{ref::exo} that it does not limit the sensitivity of current LXe detectors. 
As a condensed noble gas,
Xe is readily purified from most radioactive
impurities with one notable exception: $^{85}$Kr, with an isotopic abundance of
$^{85}$Kr/$^{\textnormal{\scriptsize nat}}$Kr$\ \sim\ $10$^{-11}$~\cite{ref::ChenCY99}.
This isotope is produced in uranium and plutonium fission and is
released into the environment in nuclear weapon tests and by nuclear reprocessing plants.

The beta decay of $^{85}$Kr with an endpoint of 687~keV and a half-life of 10.76~years
presents a serious background for the dark matter search.
Its concentration in the detector can be measured using a 
second decay mode, $^{85}$Kr($\beta$, 173~keV) $\rightarrow$
$^{85m}$Rb($\gamma$, 514~keV) $\rightarrow$ $^{85}$Rb, with a 0.454\% branching
ratio. The lifetime of the intermediate state is 1.46~$\mu$s and provides a clear
delayed coincidence signature. 

\begin{figure}[b!]
\centering
  \includegraphics*[width=0.35\textwidth]{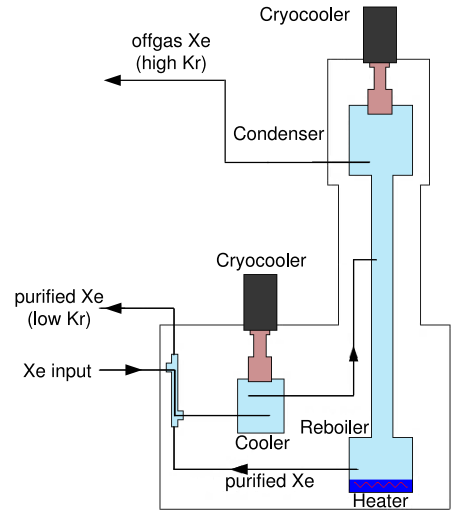}
  \caption{Schematic layout of the cryogenic distillation column used to separate krypton 
(including radioactive $^{85}$Kr) from xenon.
The XENON100 column has a height of about 3~m.}\label{fig::krcolumn}
\end{figure}

For a positive identification, events with two S1 signals and at least one
S2 signal are selected, as in some cases the two S2 signals will not be
distinguishable. The energy requirements are $20 \ \textnormal{keV}_\textnormal{\scriptsize ee}< \textnormal{S1}_1< 210 \ \textnormal{keV}_\textnormal{\scriptsize ee}$ and
$300 \ \textnormal{keV}_\textnormal{\scriptsize ee} < \textnormal{S1}_2<700 \ \textnormal{keV}_\textnormal{\scriptsize ee}$ for the first and the second S1 peak, respectively, and take into account the energy resolution. 
The timing condition $0.3 \ \mu\textnormal{s} < \Delta t <5 \ \mu$s requires a minimum separation
because two S1 peaks which are too close in time might not be identified as 
individual pulses. With these cuts, $^{85}$Kr delayed coincidence events are detected 
with $\sim$30\% acceptance, as determined by Monte Carlo simulations, and with
virtually no falsely identified events.

Commercial Xe gas contains $^\textnormal{\scriptsize nat}$Kr at the ppm level. 
Most of the gas used 
in XENON100 was processed by Spectra Gases Co.~to reduce the $^\textnormal{\scriptsize nat}$Kr content 
to the $\sim$10~ppb level~\cite{ref::krspectra}, using their cryogenic distillation
plant. During the very first XENON100 run, with a total mass of only 143~kg,  
a $^\textnormal{\scriptsize nat}$Kr level of 7~ppb was measured through the delayed coincidence analysis, 
consistent with the value provided by Spectra Gases.  
For the $^{85}$Kr-induced background to be subdominant, the fraction of 
$^\textnormal{\scriptsize nat}$Kr in Xe must be about a factor of~100 lower. 
A $^\textnormal{\scriptsize nat}$Kr/Xe ratio of 100~ppt would contribute a rate of 
$\sim \!2 \times 10^{-3}$~evts$\ \times \ $kg$^{-1}\ \times \ $keV$_\textnormal{\scriptsize ee}^{-1}\ \times \ $day$^{-1}$ 
from $^{85}$Kr~\cite{ref::xe100_mc}.

To reduce the Kr in the Xe filling XENON100, a small-scale cryogenic distillation column~\cite{ref::TayoNipponwebsite} was procured and integrated into the XENON100 system 
underground. 
The column, based on a McCabe-Thiele scheme~\cite{ref::krcolumn,ref::mccabe}, is designed to deliver a factor of~1000 in Kr reduction in a single pass, at a purification speed of 1.8~SLPM (or 0.6~kg/hour). 
A small sample of Xe gas processed with a column of similar design and analyzed by mass spectroscopy, 
was reported to have a Kr level of 3~ppt \cite{ref::krcolumn}. XENON100, however, is the first 
low background experiment using a large mass of LXe, which is sensitive to a Kr contamination at 
such ultra-low levels. 

A schematic of the XENON100 column is shown in
Fig.~\ref{fig::krcolumn}. The Xe gas is cooled using a cryocooler
before entering the column at half height. A constant thermal
gradient is kept using a heater at the bottom of the column and another
cryocooler at the top. Thanks to the different boiling
temperatures of Kr (120~K at 1~atm) and Xe (165~K) it is possible to have a
Kr-enriched mixture at the top of the column and a Kr-depleted 
one at the bottom. The Xe with a high Kr concentration is 
separated by freezing it into a gas bottle, while the Xe at the bottom is used 
to fill the detector.

After installation and an initial commissioning run of the new column, a second distillation of the full
xenon inventory was performed in Summer 2009. For the commissioning run leading to the 
first science results~\cite{ref::xe100_11d}, the 
Kr concentration was $(143 {+135 \atop -90})$~ppt (90\%~CL), as measured with the delayed coincidence method. 
This concentration agrees with the value inferred from a comparison of the measured
background spectrum with a Monte Carlo simulation \cite{ref::xe100_mc}.

A small leak in the recirculation pump before the first science run~\cite{ref::xe100run08}
led to a Kr~increase of a factor $\sim$5. This higher level did not have a large impact on the scientific reach, as demonstrated by the results~\cite{ref::xe100run08}. 
In the meantime, a lower Kr concentration, comparable to the one in~\cite{ref::xe100_11d}, was achieved by further distillation in late 2010.

\subsection{The Slow Control System and Detector Stability}

A Java-based client-and-server
system is used to monitor all relevant XENON100 parameters, such as
detector and environmental pressures and temperatures, LXe level, Xe gas recirculation rate, 
PMT voltages and currents, anode and cathode high voltage,
nitrogen~purge flow, radon-level in the shield cavity and the environment, cryostat
vacuum, etc. 
The slow control system is constantly monitored by two independent alarm servers located
in different countries. These will invoke alarms in case user intervention is required.
This system is presented in more detail in~\cite{ref::scpaper}.

\begin{figure}[tb]
  \includegraphics*[width=0.48\textwidth]{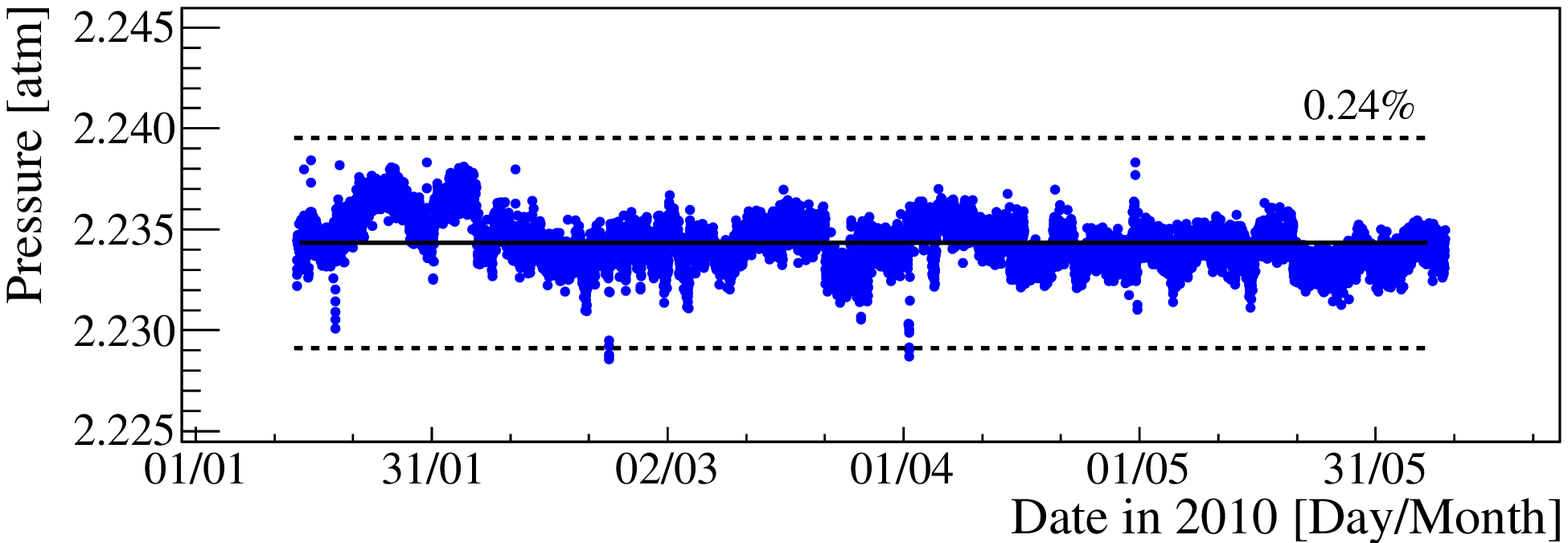}
  \includegraphics*[width=0.48\textwidth]{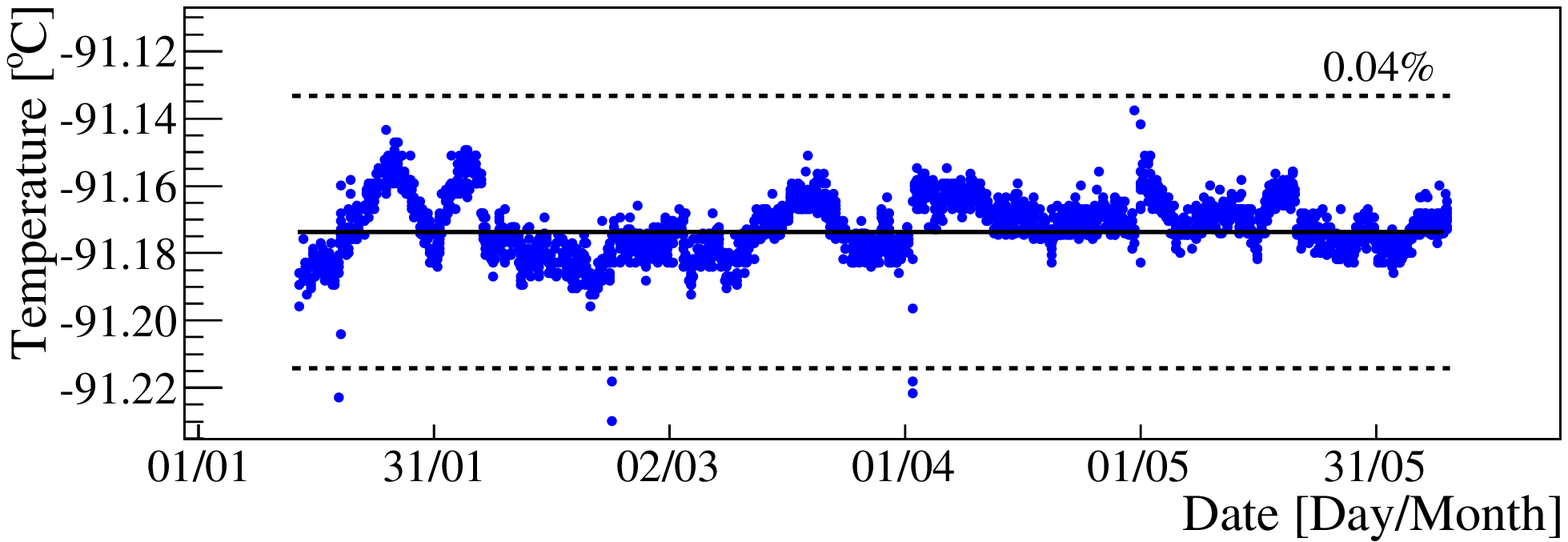}
  \caption{Long term stability of the Xe gas pressure inside the XENON100 detector (top)
and the temperature of the liquid Xe (bottom), measured over a 6~month
period. For both parameters, the fluctuations around the average are well within  
$\pm0.24$\% and $\pm0.04$\%, respectively.}\label{fig::stability}
\end{figure}

XENON100 shows excellent stability with time. Pressure and temperature, measured every 10~s with high precision, are 
stable within 0.24\% and 0.04\%, respectively. Fig.~\ref{fig::stability} shows
the evolution of these parameters for the science data reported in~\cite{ref::xe100run08}, 
covering a period of $\sim$6~months.

The size of the S2 signal is directly related to the pressure $P$ (in bar) 
of the Xe gas in the detector \cite{ref::bolozdynya1999,ref::freitas2010}: An electron extracted
into the gas phase by a uniform electric field~$E$ (in kV/cm) generates
$n_{\textnormal{\scriptsize ph}}$ photons,
\begin{equation}\label{eq::s2}
n_{\textnormal{\scriptsize ph}} \propto \left(\frac{E}{P} - 1.0 \right) P x
\textnormal{,}
\end{equation}
where $x$ is the gas gap between LXe surface and the anode (in cm). At the XENON100
operating conditions, pressure fluctuations of $0.24$\% lead to negligible S2~signal
fluctuations of \mbox{$<0.05$\%}.

\subsection{The Data Acquisition System}\label{sec::daq}

\begin{figure}[tb]
\includegraphics*[width=0.48\textwidth]{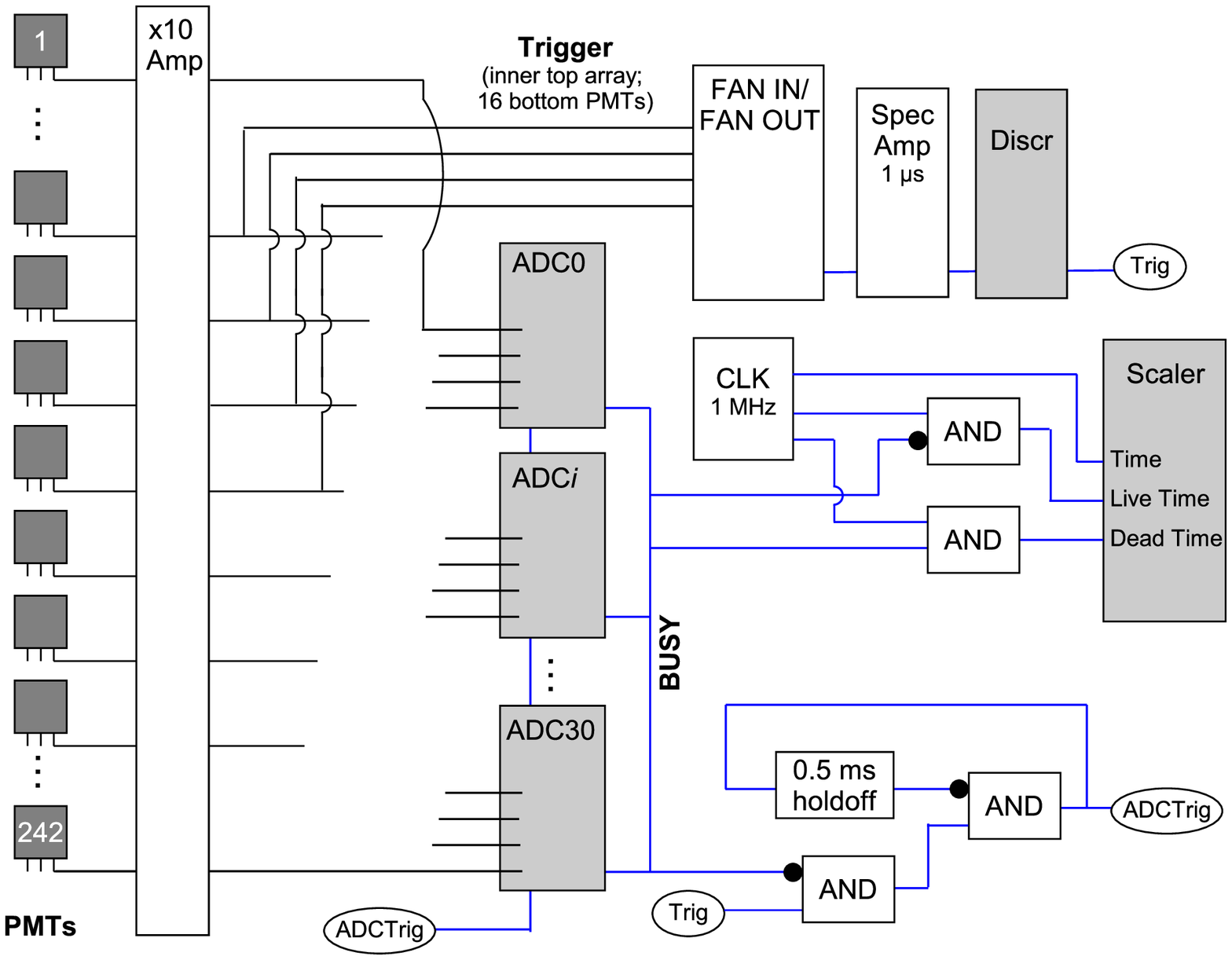}
\caption{DAQ schematic of XENON100 for the dark matter search. All 242~PMTs are digitized at
100~MHz sampling rate with Flash ADCs. A hardware trigger 
is generated using a sum of 68~top and 16~bottom PMTs. The ellipses in the figure indicate
connection lines.}\label{fig::daq}
\end{figure}

The XENON100 data acquisition (DAQ) system generates the trigger, 
digitizes the waveforms of the 242~PMTs, and stores the data in an indexed file format.
The general DAQ~layout is shown in Fig.~\ref{fig::daq}.

The PMT signals are amplified by a
factor~10 using commercial Phillips~776 NIM amplifiers 
and then digitized by CAEN VME V1724
Flash ADCs with 10~ns sampling period, 14~bit resolution, 2.25~V full scale, and 40~MHz
bandwidth. At the typical background rate of $\sim$1~Hz, a deadtime-less measurement 
is possible, since the ADC has a circular buffer 
with 512~kB memory per channel. An on-board FPGA allows to operate the ADCs in a mode where
only parts of the waveform around a peak above a certain threshold are
digitized (zero-length-encoding). The position of the peaks within the waveform is stored. 
For XENON100, this threshold is 4~mV for each channel, corresponding to about 0.3~photoelectrons. 
Since large parts of an event's waveform consist only of
baseline, this leads to a decrease in data size by more than a factor of~10.

For high energy $\gamma$-sources, this mode allows calibration rates that are more than 
one order of magnitude larger
than the background rate. The rate is limited to about 30~Hz by the tail of 
large S2 signals that contains many small peaks exceeding the digitization
threshold. For low energy sources, the rate can be higher.

The trigger is generated using the signals of the 68~inner PMTs of the top
array plus 16~PMTs from the bottom center, summed with linear Fan-In/Fan-Out modules. 
It is amplified, integrated ($\tau=1$~$\mu$s), and shaped with a spectroscopy
amplifier (ORTEC~450) in order to be able to trigger on very small S2 signals
that consist of $\sim$300~PE (corresponding to $\sim$15~ionization electrons), spread within 1~$\mu$s.
A low threshold discriminator generates the digital trigger signal
which is distributed simultaneously to the 31~ADCs. 
A square-wave voltage pulse of 1~$\mu$s width fed into the trigger signal line verified that the
trigger threshold was 100\% at a pulse height of 24~mV, which corresponds to 150~PE. As not all PMTs of the TPC contribute to the trigger but still detect part of the S2~light, this number has to be corrected by the average fraction of S2 signal seen by the triggering PMTs. It was measured to be $52\%$ of the full S2
signal, leading to a S2~threshold of 290~PE. 
This number was confirmed by a direct measurement of the trigger threshold: calibration waveforms of 400~$\mu$s length were
acquired together with the digitized logical trigger signal which was recorded by an unused ADC channel. This allowed for a direct comparison
of the spectrum of S2 signals that generated a trigger (identified by the simultaneously stored logical trigger signal) to the full S2 spectrum (S2 signals without condition on the trigger signal) and to derive the trigger threshold.
Additional tests with lowered discriminator
thresholds (where many events are noise triggered) also indicate that the trigger
threshold is $>99\%$ above 300~PE.

The trigger based on the analog sum is used for standard detector
operation, however, there is also the possibility to trigger on a majority signal, i.e.,
when a certain number of PMTs exceeds a threshold. This mode is used for
measurements without drift field when no S2 signal is generated and to take data triggering on the active veto. The threshold is
higher in this configuration due to the short time coincidence window (10~ns) of the majority signal.

After the science run which lead to the results published in~\cite{ref::xe100run08}, 
the trigger has been modified in order to lower the threshold. Every ADC generates 
a 125~mV square-wave output signal for every channel exceeding a pre-defined signal 
height, which is set to $\sim$0.5~PE and leads to a strong amplification of small 
signals. The sum of the analog signals is integrated using a spectroscopy amplifier 
with a time constant of 1~$\mu$s and fed into a low threshold discriminator. The 
improved trigger threshold is $>$99\% above S2~$\sim$150~PE, as derived from a 
dedicated measurement in which the trigger signal was recorded.

In order not to miss any waveform information, regardless whether the trigger is generated by
a S1 or a S2 signal, the trigger is placed in the middle of the event window of
400~$\mu$s length, more than twice the maximum electron drift time
of 176~$\mu$s at 0.53~kV/cm drift field. A 500~$\mu$s trigger hold-off is
applied after a trigger in order to avoid event overlaps. 
Additionally, a trigger is blocked when the ADCs are busy or when a high energy veto
is applied. 

The high energy veto is implemented to reduce the amount of data during calibration
at the low energies of interest for the dark matter search. 
The idea of the veto is to inhibit events which would be triggered by the S1 signal as this 
is possible only for high energy events. 
S1 peaks are identified by selecting peaks with a narrow width (to distinguish narrow S1 from wide S2 peaks) 
by shaping and differentiating the signal in a spectroscopy amplifier. A threshold is then 
set on the size of the 
S1 peak and events with a S1 signal above the threshold are rejected. In this case 
further triggers are inhibited for the next 500~$\mu s$ in order to prevent 
triggers generated by subsequent S2~peaks. Because of the S2 amplification these 
would be even larger than the S1 signal generating the veto.

A VME scaler is used to measure the detector live time and dead time using an
external 1~MHz clock and the BUSY outputs of the ADCs. The effect of the trigger
holdoff is taken into account separately. The deadtime during science data taking
is about~1\%. 

The time of every accepted trigger is recorded with microsecond resolution.
For easier access to the raw waveforms of a particular event, the data are stored
in an indexed file format that can
be compressed further using standard compression tools during data taking. 
The extraction of physical
parameters from the waveforms is done offline on a computing cluster separated
from the DAQ system (see~Sect.~\ref{sec::rawdata}).

%% file: rawdata.tex
\section{Low-level Analysis}\label{sec::zeroorder}

In this section, we focus on how physical quantities are obtained from the raw 
waveforms recorded by the DAQ system,
which can then be used for data analysis.

\subsection{PMT Gain Calibration}\label{sec::pmtcal}

\begin{figure}[b!]
  \centering
  \includegraphics*[width=0.48\textwidth]{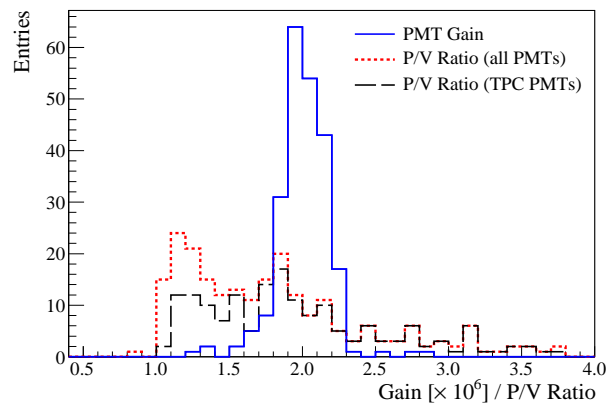}
  \caption{Distribution of gain and peak-to-valley (P/V) ratio of the PMTs in XENON100. The gain is set to be around $2.0 \times 10^6$. Tubes with a better single photoelectron response (higher peak-to-valley ratio) are selected for the TPC.}\label{fig::gaindistr}
\end{figure}

The bias voltage for each of the 242~PMTs is set individually and is
chosen such that its gain is close to $2.0 \times 10^6$ (see Fig.~\ref{fig::gaindistr}). 
Typical deviations from
this gain value are within $\pm$10\% and are due to voltage constraints for a given PMT. The
gains are determined by stimulating single photoelectron emission from the
photocathode using light pulses from a blue LED ($\lambda=470$~nm), driven by a pulse generator.
The light is generated outside the detector and fed into the system via two
optical fibers, one for the TPC and one for the active xenon veto. Above the diving bell,
these two main fibers split 
into four (six for the veto) in order to distribute the light more uniformly. 
The light level is chosen
such that in $<$5\% of the LED triggers, a PMT shows a photoelectron signal in the time window 
considered for the analysis. When this is not the case, the window is adjusted in the analysis 
procedure in order to increase or decrease the photoelectron probability. 
This procedure allows for the calibration of all 242~PMTs in one run.
The light level is sufficiently low to avoid any relevant contamination from two coincident
photoelectrons. A pulser frequency of only 100~Hz ensures that the PMT bases are fully
charged between two light pulses.

\begin{figure}[tb]
  \centering
  \includegraphics*[width=0.3\textwidth]{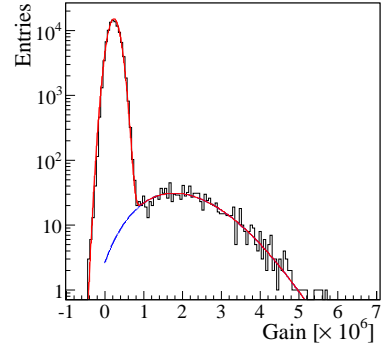}
  \caption{Single photoelectron spectrum of a typical PMT installed in XENON100. The measured
signal (in photoelectrons) has been already converted into the PMT gain.
The position of the single photoelectron peak is obtained in a simultaneous fit
together with the much higher noise peak close to the
origin. This particular PMT has a gain of $2.06\times 10^6$, and a peak-to-valley
ratio of 1.54.}\label{fig::spespectrum}
\end{figure}

The position of the single photoelectron peak in the pulse area 
spectrum (see~Fig.~\ref{fig::spespectrum}) is
directly proportional to the gain of the tube.
It is determined by a simultaneous fit of the single photoelectron peak and the noise peak.
The latter is described by a Gaussian function whereas the single photoelectron peak is given by the 
continuous distribution~\cite{ref::alexa}
\begin{equation}
y(x) = \frac{\mu^x \exp(-\mu)}{\Gamma(x+1)}\textnormal{,}
\end{equation}
where $y(x)$ is the number of counts in the spectrum, $\mu$ is the mean of a Poisson distribution, and
$\Gamma$ is the Gamma function. The fit quality is ensured by checking the $\chi^2$-value and by visual
inspection of all fits.
For monitoring purposes, the individual PMT gains are measured once a week: They show
fluctuations within $\pm$2.0\% (1$\sigma$), in 
\onecolumn
\begin{figure}[p]
  \centering
  \includegraphics*[width=0.6\textwidth]{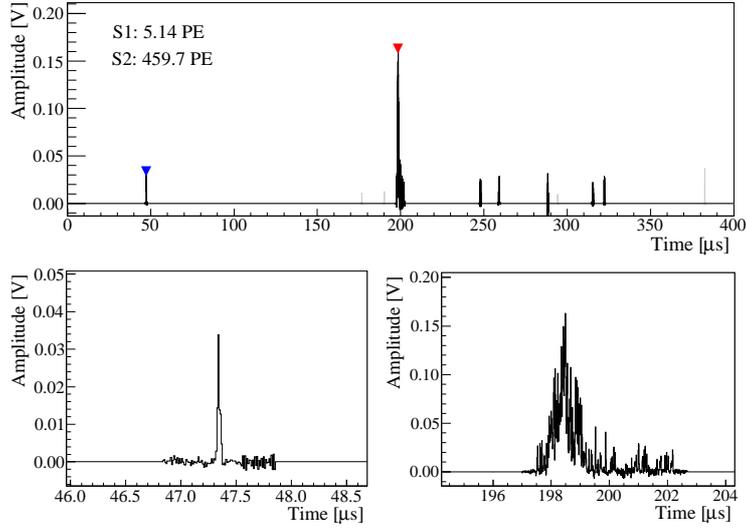}
  \caption{Example of a low-energy event from background data: The top figure
shows the full waveform (400~$\mu$s), which is the sum of the 178 PMTs of the TPC. At
$\sim$47~$\mu$s is the S1 peak (blue marker) of 5.1~photoelectrons (PE), the S2 peak at $\sim$200~$\mu$s
(red marker) consists of 460~PE. No position-dependent corrections have been applied to these values. 
The time difference (=drift time) between the peaks is
151~$\mu$s. The small structures after the S2 peak are S2 signals from
single electrons extracted into the gas phase. With a size of $<40$~PE they are
clearly distinguishable from the main S2 peak. 
 Closer views of the S1 and the S2 peak are shown on the bottom left and
right, respectively.
}\label{fig::waveform}
\end{figure}

\begin{figure}[p]
  \centering
  \includegraphics*[width=0.65\textwidth]{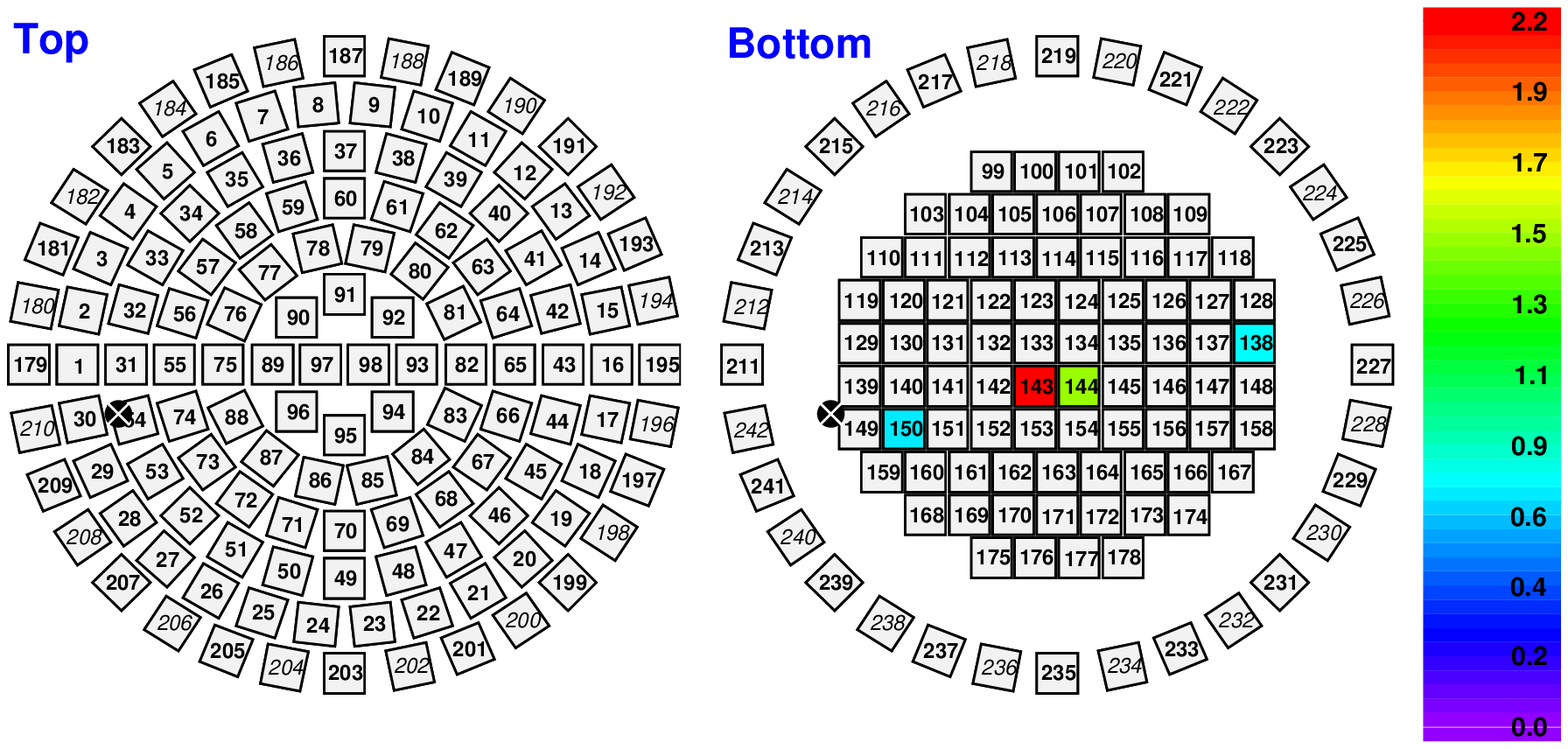}
  \includegraphics*[width=0.65\textwidth]{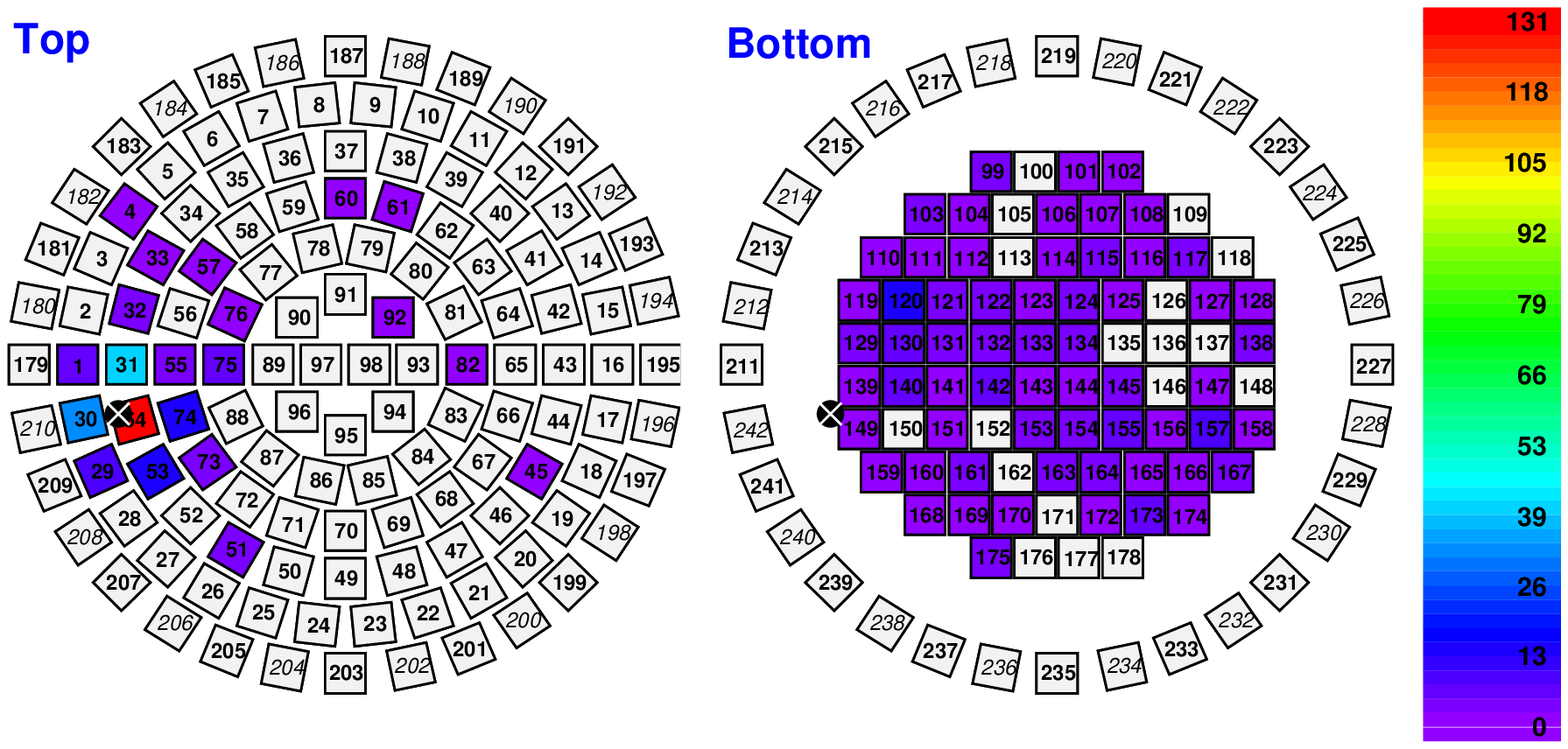}
  \caption{PMT hit pattern of the event displayed in Fig.~\ref{fig::waveform}.
Numbers indicate individual PMTs, PMTs 179-242 are in the active veto.
The color code is proportional to the signal (in PE) seen by the respective PMT.
(Top) S1 hit pattern. Only 4~bottom PMTs see the S1 peak due
to the small energy of the event. (Bottom) S2 hit pattern. The hit pattern on the top 
array (bottom left) is used
for the $(x,y)$ position reconstruction. In both figures, the marker indicates the
$xy$-position reconstructed by the neural network algorithm. }
\label{fig::hitpattern}
\end{figure}
\twocolumn

\noindent agreement with the overall
uncertainty of the gain determination and much lower than the S1 resolution of
the detector. The average gain during the science run was stable within $\pm 1$\%.

\subsection{Peak Identification}\label{sec::rawdata} 

The raw data processor, a ROOT \cite{ref::root} based C++ program, is used to
derive physical quantities from the waveforms. An
event consists of the traces of all 242~PMTs of typically 400~$\mu$s length. Since the 
PMT trace between peaks is not recorded (zero-length-encoding, see Sect.~\ref{sec::daq}), 
a waveform consists of several digitized sub-waveforms together with their respective 
position in time.

The processing is done in several steps: First, the waveforms are reconstructed
from the raw data. The baseline is calculated and subtracted individually for each
PMT and for each recorded waveform section. The amplitudes are converted to
voltage (see~Fig.~\ref{fig::waveform}). For the peak finding, the waveforms of
the 178 inner PMTs (ignoring 3-4 PMTs that exhibit an increased dark count rate)
are summed up. A digital low-pass filter (32~tap filter with a cut-off frequency of
3~MHz) is applied in order to remove high
frequencies from the signal to facilitate the determination of the extent of
a peak. First, the program searches for S2 peak candidates on the filtered 
waveform by finding waveform intervals $I_\textnormal{\scriptsize S2}$ that 
exceed a threshold of 10~mV
for at least 600~ns. In order to be considered as S2 peak candidate, the average
signal in the 210~ns before and after $I_\textnormal{\scriptsize S2}$ has to be 
lower than 25\% of the maximum peak height within $I_\textnormal{\scriptsize S2}$.
Due to multiple scattering, afterpulses, and small S2 signals from single electrons, 
an interval $I_\textnormal{\scriptsize S2}$ might contain several S2 peak
candidates. These are identified recursively on the filtered waveform by searching for 
clear minima close to the baseline between the peaks (bin content below 0.1\% of the peak maximum), or for a sign-change in the slope
of the waveform. For all peak candidates with a FWHM $>350$~ns, parameters such as
the position in the waveform, area and width, PMT coincidence level, etc.~are 
calculated. The peak integration is performed over the full extent of the peak as determined above.

The corresponding S1 signal has to be found preceding the S2~signal in time. The much 
larger S2 signals as well as afterpulsing and S2s from single electrons make an
identification of S1 peaks after an S2 signal difficult. Therefore, the algorithm does not
attempt to identify S1 signals after the first S2 peak exceeding 300~PE.
The S1~peak finder scans the un-filtered sum waveform for peaks exceeding a threshold
of 3~mV ($\sim$0.33~PE). The waveform regions before and after the peak candidate are taken into account 
in order to decide whether it is a real peak, and the fluctuations around the baseline
are examined in order to reject electronic noise. 
The detection efficiency for small S1 peaks is very high: $>$80\% for single photoelectrons, $>$95\% for double
photoelectrons, and $>$99\% for 3~photoelectrons.
A similar analysis is done on the sum of the active veto PMTs.

For every detected peak candidate, peak properties such as height, width, area,
position in the waveform, PMT coincidence level, etc.~are determined using the
information of all available PMTs. All information is stored for every peak candidate. The
identification of valid single-scatter event candidates is done based on these parameters at a
later stage in the offline analysis. For example, a valid S1 peak 
is typically required to be seen on at least two PMTs simultaneously and its width (in ADC samples) 
must be above a certain threshold in order to reject electronic noise. 

The gains of the PMTs are measured and monitored independently (see~Sect.~\ref{sec::pmtcal}).
Using the gain values for each PMT, averaged
over a few months, the peak areas are converted into photoelectrons.

\subsection{3D Vertex Reconstruction}\label{sec::posrec}

One advantage of the TPC technique for dark matter searches is the possibility
to determine all three coordinates of an interaction vertex in the target volume on
an event-by-event basis, as this allows for position-dependent signal
corrections and fiducial volume cuts for background suppression.

In a uniform electric drift field, the electron drift velocity is constant (1.73~mm/$\mu$s under the present operation conditions)
and the $z$-coordinate is determined from the time difference $\Delta t = t_{S2} - t_{S1}$ between the prompt 
S1 and the delayed S2 signal. $t_{S1}$ and $t_{S2}$ are determined at the maxima of the pulses. 
From the maximum drift time and the known TPC length
(or via the independently measured drift velocity \cite{ref::miller1968}) this can 
be converted to the space coordinate $z$. The $z$-position resolution of XENON100 is 0.3~mm (1$\sigma$) as
inferred from events in background data at a well known position near the top liquid layer, the gate grid, or the cathode. Because of the finite 
width of the the S2 signal, two S2 pulses can only be separated if they are more than 3~mm in $z$ apart.

The determination of the $(x,y)$ position exploits that the secondary S2 signal
from the charge cloud is generated at a very localized spot right above the liquid-gas 
interface. 
This leads to a highly clustered S2 signal on the array of top PMTs, which is detected using the fine granularity
of the 1"$\times$1" PMTs (see Fig.~\ref{fig::hitpattern}). Three different position reconstruction 
algorithms have been developed to obtain the $(x,y)$ position from a comparison of the measured
top array PMT hit pattern with the one generated by a Monte Carlo simulation. The presence of
non-functional PMTs has been taken into account in all algorithms.

The $\chi^2$-algorithm compares the observed PMT hit pattern with the ones from the simulation 
and finds the $(x,y)$ position by minimizing the $\chi^2$-value. An advantage of this method
is that the $\chi^2$ can be used to quantify the quality of the reconstructed position.

Another approach is utilized by the support vector machine algorithm (SVM) \cite{ref::SVM}: This mathematical 
procedure is based on training samples from the Monte Carlo simulation, using the signal proportion on
each PMT in the top array for a given $(x,y)$ position, to create a base of vectors
in a projected multidimensional space. 
The $(x,y)$ position of an interaction can then be found by solving a sum of scalar products 
between the vector corresponding to the measured input data and the base vectors defined in the training
of the algorithm. The SVM algorithm was used for the analysis presented in Ref.~\cite{ref::xe100_11d}.

\begin{figure}[b!]
  \centering
  \includegraphics*[width=0.38\textwidth]{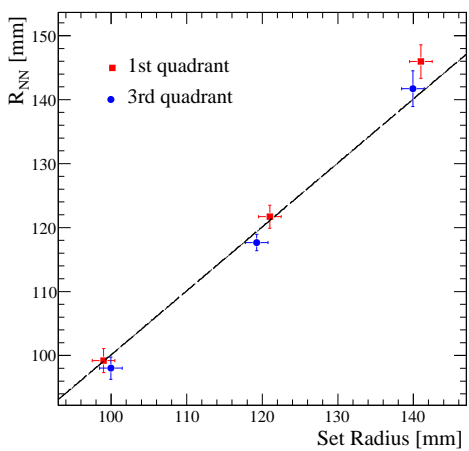}
  \caption{Set position of the S2 spot from a collimated $^{57}$Co source vs.~the position 
reconstructed by the neural network algorithm. The collimator was moved across the detector, 
hence datapoints exist for the first and the third $(x,y)$ quadrant. The measurement focused 
on large radii, where position reconstruction is most crucial because of the fiducial volume
cut. Within the uncertainties, set positions and reconstructed positions agree. The diagonal line shows where
set and measured radius coincide.}
\label{fig::collimator}
\end{figure}

The third algorithm uses a neural network (NN) \cite{ref::stuttgartNN}: The network consists of units (neurons) with
several inputs from other units, and one output whose value is calculated from the inputs. Here, a
feed-forward multilayer perceptron is employed.  
98~inputs (the PMTs on the top array) lead to the output position vector $(x,y)$. They are 
connected by one~hidden 
layer with 30~neurons. The network must be trained (backpropagation rule) 
using Monte Carlo data before
it can be used to derive the position corresponding to measured hit patterns. More details can 
be found in Ref.~\cite{ref::xe10instrument}.

All three algorithms give results consistent with each other and with the Monte Carlo simulation for radii
$r<142$~mm. A $(x,y)$ resolution of 
$<3$~mm ($1\sigma$) was measured with a collimated $^{57}$Co source located above the TPC at several $r_i$
while the LXe veto above the diving bell was not filled.
This result, shown in Fig.\ref{fig::collimator}, 
is dominated by the point spread due to the finite size of the collimator opening. The expected 
resolution based on Monte Carlo data is better than 2~mm. For the science results published in~\cite{ref::xe100run08}, the NN algorithm was used
as it shows the most homogeneous response and better agreement with the expectation from a Monte Carlo simulation. The other two algorithms were used to cross check the position obtained with NN 
and for data quality cuts.

%% file: calibration.tex
\section{Detector Optimization and Characterization}\label{sec::calibration}

In this section we describe measurements and analyses
that have been performed with XENON100 during commissioning in order to characterize the 
detector response. In particular, we describe the position dependent corrections
which are applied to the data.

\subsection{Calibration and Calibration Sources}

In order to characterize the detector, calibration sources can be inserted in
the XENON100 shield through a copper tube which is wound around the 
cryostat (see Fig.~\ref{fig::xe100shield}). 
While vertical source position is restricted to
the TPC center, it can be placed at all azimuthal angles.

The sources $^{137}$Cs, $^{57}$Co, $^{60}$Co, and $^{232}$Th are used for
gamma calibrations. The electronic recoil band in $\log_{10}(\textrm{S2/S1})$ vs~energy space 
defines the region of background events from $\beta$- and $\gamma$-particles. It is 
measured using the low energy Compton tail of high-energy $\gamma$-sources such as $^{60}$Co and
$^{232}$Th. 
The response to single scatter nuclear recoils, which is the expected signature of a WIMP, 
is measured with an $^{241}$AmBe
$(\alpha,n)$-source, shielded by 10~cm of lead in order to suppress
the contribution from its high energy gamma rays (4.4~MeV). Besides the definition 
of the nuclear recoil band and a benchmark
WIMP search region, this calibration provides additional gamma lines from inelastic
scattering as well as from xenon or fluorine activation at 40~keV ($^{129}$Xe), 80~keV ($^{131}$Xe),
110~keV ($^{19}$F in PTFE), 164~keV ($^{131m}$Xe), 197~keV ($^{19}$F), and
236~keV ($^{129m}$Xe).

\subsection{Detector Leveling and S2 Optimization}\label{sec::level}

Even though the position reconstruction capabilities of the TPC allow to correct for
spatial detector anisotropies, it is beneficial to minimize them in the first place.
The size of the S2 signal depends on the width $x$ of the gas gap
between the liquid-gas interface and the anode, see Eq.~(\ref{eq::s2}). For an optimal detector response, 
the detector therefore has to be leveled to get a uniform proportional scintillation S2~signal, 
and the liquid
gas interface has to be adjusted at the optimal height to optimize the S2 resolution.
Leveling of the detector within the closed shield was achieved by adjusting 
set screws in two of the three support legs which can be accessed from the outside of the shield.
 
\begin{figure}[tb]
  \centering
  \includegraphics*[width=0.4\textwidth]{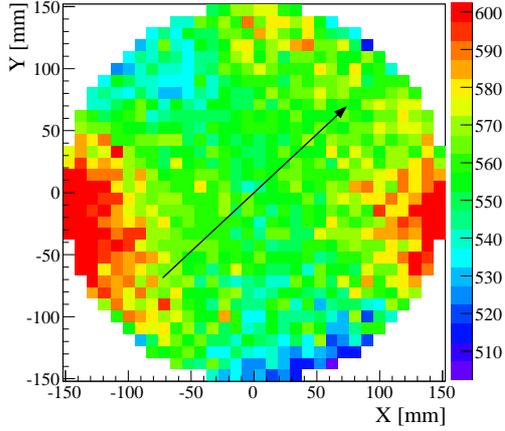}
  \caption{Measured width $\Delta w$ (FWHM in ns) of the S2 signal of a backscatter peak in XENON100 background data. Only the topmost layer of LXe is selected. The variation due to warping of the top meshes is maximal at the edges. The arrow gives the maximum overall variation across the whole TPC, which is minimized in the leveling procedure.
}\label{fig::level}
\end{figure}

For a given calibration peak, the width $w$ of the S2 signal (FWHM in ADC samples) is directly proportional
to $x$, hence the variation $\Delta w$ of the S2 width along a line~$\ell$ across the LXe surface was used as a measure for the leveling (see Fig.~\ref{fig::level}). 
The leveling was done iteratively using a $^{137}$Cs calibration source at several position around the 
detector. The variation $\Delta w/\ell$ was inferred by fitting a plane to the measured S2 width  
using only data from the topmost layer of LXe ($\sim$3~mm). It was then reduced accordingly by tilting the detector.
After the leveling procedure the value across the line of maximum variation was
$\textnormal{max}(\Delta w/\ell) = 0.3$~ns/cm for the remaining S2~variation. It
was measured with the dominant backscattering peak in homogeniously distributed background data,
again using only the topmost layer of LXe.

These analyses use only basic data selection cuts and do not take into account events 
close to the borders of the TPC, where S2 width variations from mesh warping are larger. Locally, these effects can cause much larger variations of up to 50~ns.
The S2~pulses are additionally widened due to longitudinal dispersion of the electron cloud drifting in the LXe.
This is the dominant effect for S2~width variations, leading to widths of $\sim$1~$\mu$s for events occuring close to the cathode.

The S2 response was optimized further in the leveled configuration, again using a 
radially uniform distributed $^{137}$Cs 
source, by varying the overall liquid xenon level until the 
best resolution of the full absorption S2~peak was found (see Sect.~\ref{sec::ces}).
The height of the liquid was monitored with capacitor levelmeters and 
the best performance was achieved with a LXe level of 2.5~mm above the gate grid.

\subsection{Position Dependence of the Charge Signal}\label{sec::s2corr}

After detector leveling (see Sect.~\ref{sec::level}), two independent effects 
remain that have an impact on the size of the proportional scintillation signal S2 from the charge:
The first is due to to absorption of electrons as they drift (finite electron lifetime), 
leading to a $z$-dependent correction.
The second is due to a reduced S2 light collection efficiency at large radii,
non-functional PMTs, quantum efficiency differences between neighboring PMTs, 
as well as non-uniformities in the proportional scintillation gap. 
This leads to a S2 correction that depends on the ($x,y$)-position of the S2 signal.

\begin{figure}[b!]
  \centering
  \includegraphics*[width=0.48\textwidth]{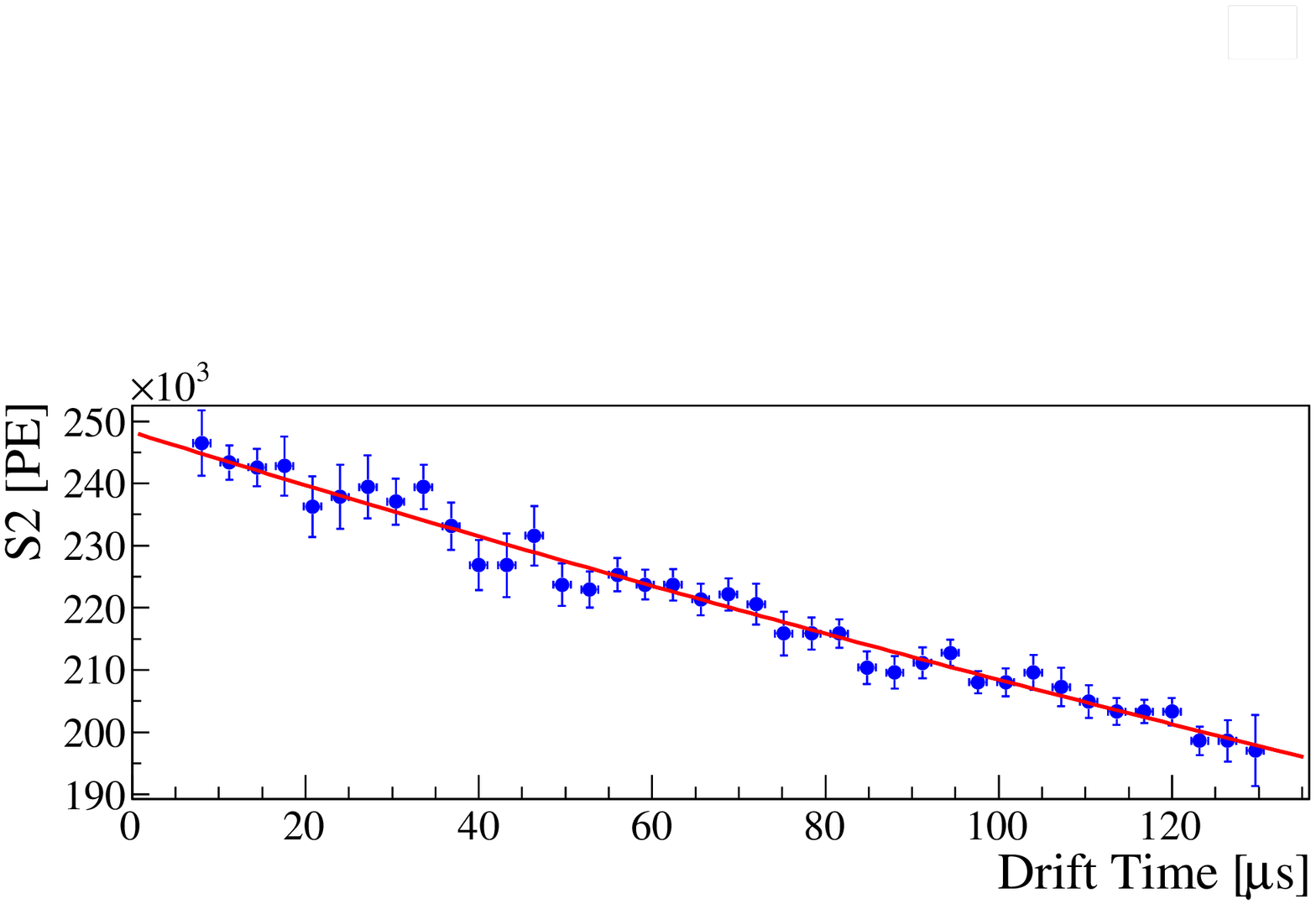}
  \caption{The charge loss due to impurities in the LXe is determined regularly by measuring the S2 signal of the $^{137}$Cs full absorption peak vs the electron drift time. An exponential fit to the data yields the electron lifetime $\tau_e$, which is 556~$\mu$s in this example. The used drift time range is limited to an interval where background and electric field inhomogeneities (Sect.~\ref{sec::efieldCorr}) are irrelevant.
}\label{fig::elife}
\end{figure}

As discussed in Sect.~\ref{sec::lxepurification}, the amount of free electrons generated in an
interaction is reduced by absorption on electronegative impurities (mainly oxygen) in the LXe while the
electrons are drifted upwards in the drift field. For a given particle energy $E$, this
is described by the exponential relation
\begin{equation}
\textnormal{S2}(E) = \textnormal{S2}_0(E) \ \exp\left(-\frac{\Delta t}{\tau_e}\right), 
\end{equation}
where S2 is the observed secondary signal and S2$_0$ is the signal that would have been 
observed in the absence of drift-time dependent losses.
$\Delta t$ is the
drift time, which is the time difference between the prompt S1 and the delayed S2 signal.
$\tau_e$ is the electron lifetime
describing charge losses. Since $\tau_e$ is increasing with improved LXe purity (see Fig.~\ref{fig::chargeevolution} on page~\pageref{fig::chargeevolution}), 
it is measured 
regularly using a $^{137}$Cs source, see Fig.~\ref{fig::elife}. As the changes of $\tau_e$ with time are small 
($\sim$1~$\mu$s/day) a calibration measurement once to twice a week is sufficient. 
Because the $\tau_e$ trend shows a linear behavior the data is corrected using a linear fit to all $\tau_e$ measurements to reduce the impact of statistical fluctuations.

During the science run leading to the 
results published in Ref.~\cite{ref::xe100run08}, $\tau_e$ was increasing from 230~$\mu$s to 380~$\mu$s.
The drift time correction is largest for events
close to the cathode. With $\tau_e>230 \ \mu$s and a maximal drift time of 176~$\mu$s, this
leads to a maximum correction smaller than a factor~2. In the meantime, even higher $\tau_e$ values have been reached as shown in Fig.~\ref{fig::elife}.

\begin{figure}[tb]
  \centering
  \includegraphics*[width=0.4\textwidth]{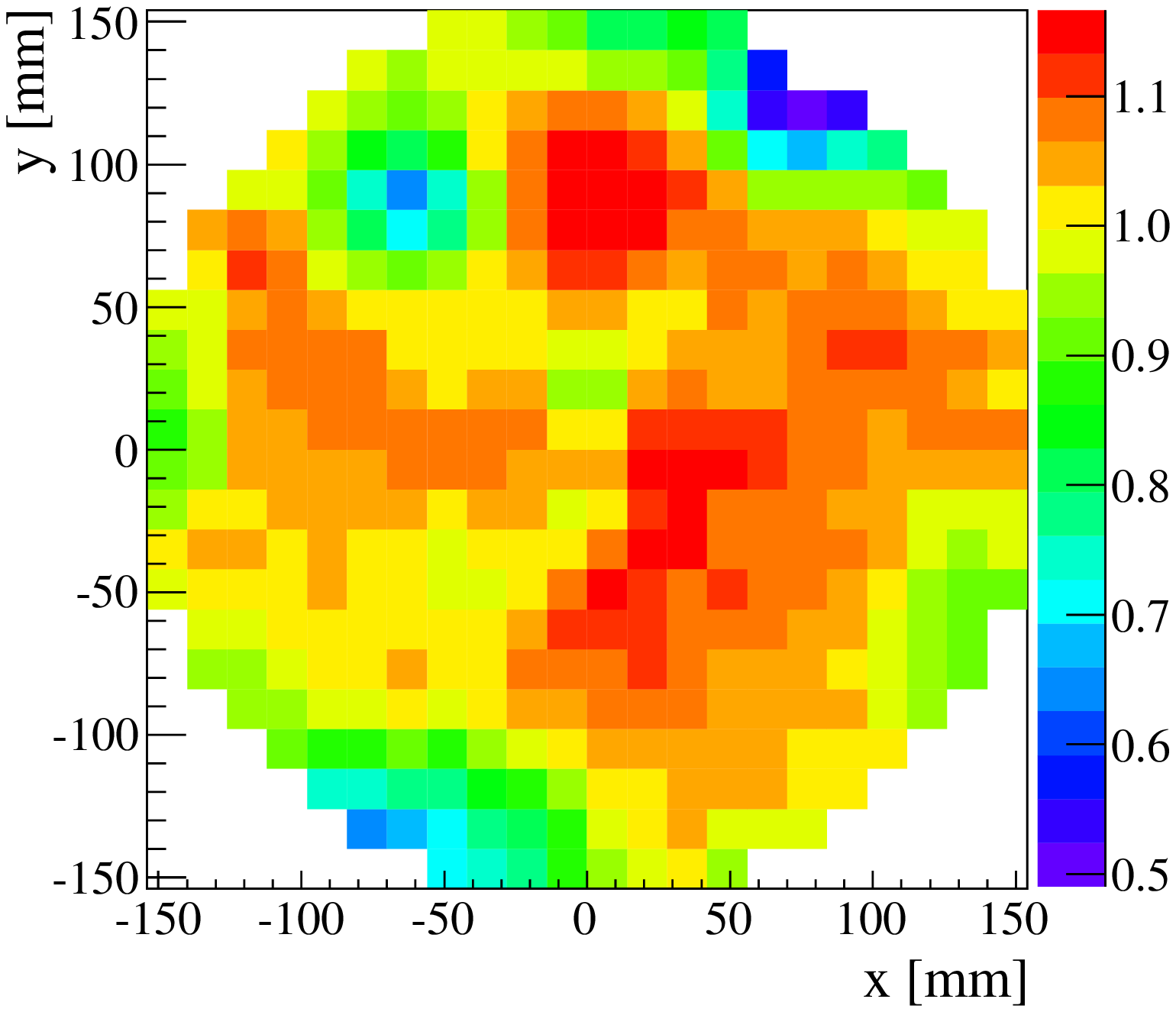}
  \includegraphics*[width=0.4\textwidth]{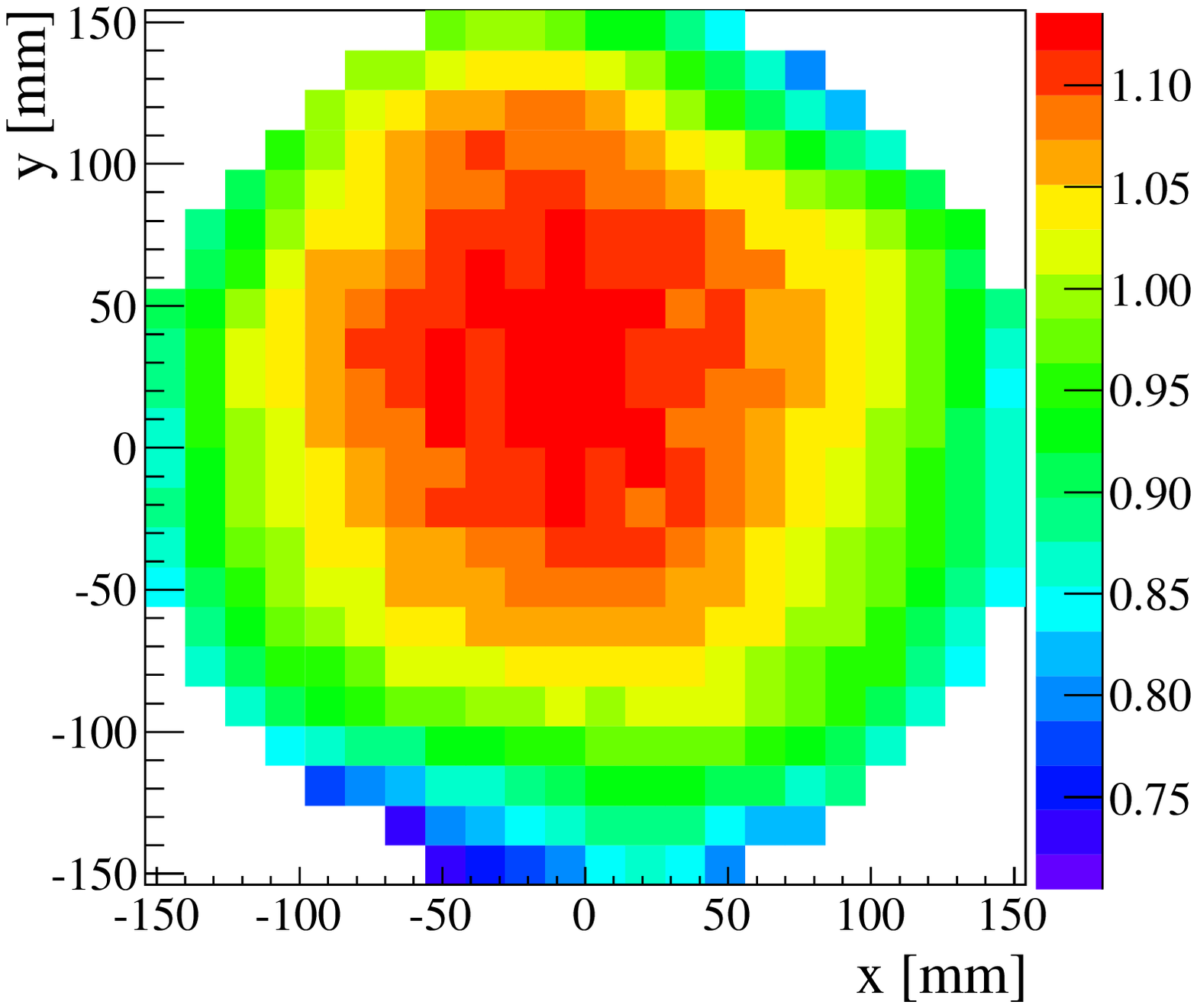}
  \caption{S2 response of the top and the bottom PMT array, measured with the 40~keV line from inelastic 
neutron scattering on $^{129}$Xe. Shown color-coded is the relative change compared to the mean. The decrease at large radii is due to a reduced S2 light collection efficiency. The top array shows more fluctuations due to non-working PMTs (e.g.~at $x\sim-50$~mm, $y\sim100$~mm).
}\label{fig::s2corr}
\end{figure}

The solid angle covered by the PMT arrays decreases for S2 events generated close to the 
PTFE walls, leading to a lower S2 light collection efficiency and therefore to a smaller
observed S2 signal. Other effects such as non-functional PMTs also cause spatial signal variations. 
The correction due to these effects was determined in the same way as
described in Sect.~\ref{sec::lce} for the S1 light collection efficiency, using
$^{137}$Cs as well as the 40~keV and 164~keV gamma lines from inelastic neutron scattering and
neutron activation, respectively. The corrections obtained from the three lines agree
within the uncertainties.

Fig.~\ref{fig::s2corr} shows the S2 signal from the 40~keV line as seen by the 
top and the bottom PMT array, respectively. For all typical fiducial volumes the top PMTs record about 55\% of the S2 signal, which is on average $\sim$8600~PE for the 40~keV line summed over all PMTs and after corrections. There is almost
no spatial anisotropy on the bottom besides the decrease towards the edge which is due to the decrease in S2~light collection efficiency. The impact of non-working PMTs 
and mesh warping is irrelevant here since the large distance of $>$30~cm to the proportional 
scintillation region spreads the S2~light over the full
array. For a fiducial mass of 48~kg ($r<141$~mm), the maximum correction is about 15\% with an RMS of 7.0\%. 
On the top array, spatial S2 variations are larger, which is mostly due to regions of reduced S2 
sensitivity where individual PMTs are not working. 
Locally, this leads to larger corrections. However, the RMS value of 8.8\% is only slightly higher
than on the bottom array. The signal on both arrays is corrected independently, and only the S2 signal on the bottom array is used for S2/S1 discrimination in the analysis presented in~\cite{ref::xe100run08}.

\subsection{Light Collection Efficiency and Light Yield}\label{sec::lce}

For a given energy deposition, 
the amount of light that is actually measured by the detector depends on the position of
the interaction in the TPC, since solid angle effects, reflectivity, Rayleigh scattering
length, transmission
of the meshes, etc.~affect the light collection of the PMT arrays. Hence,
a 3-dimensional map of the light collection efficiency is mandatory to correct the data.
A detailed comparison of $^{137}$Cs data taken at
different positions around the detector confirmed that
the TPC response is indeed axial-symmetric as designed. Hence it is sufficient to
obtain a $(r,z)$ correction map instead of a $(x,y,z)$ map.   

\begin{figure}[tb]
  \includegraphics*[width=0.47\textwidth]{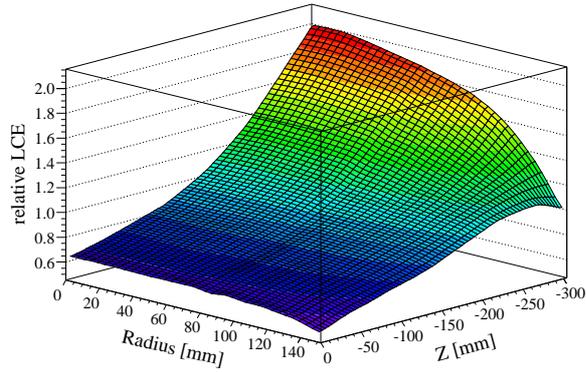}
  \caption{Correction map for the light collection efficiency (LCE) obtained from the 40~keV 
line: The vertical axis shows the value to correct a measured S1 (light) 
signal at a given $(r,z)$ position. $z=0$~mm denotes the top of the TPC. The measured S1
signals are divided by the values from this map.}\label{fig::lce}
\end{figure}

Three different sources were used to infer the correction map:
An external $^{137}$Cs source (662~keV) taken at 3 positions around the detector,
the 40~keV line from inelastic scattering of neutrons on $^{129}$Xe (one source
position), and the 164~keV line from neutron-activated $^{131m}$Xe with a uniform
distribution inside the TPC. 
For each of the sources, the light yield was determined in $(r,z)$ bins.
The bin size was decreased at larger $r$ where the light collection efficiency is
expected to fall off stronger while increased statistics from the $^{137}$Cs and $^{241}$AmBe
sources allowed for a finer binning. The light collection efficiency varies by 
a factor $\sim$3 across the TPC, with the largest value in the center, right above the
bottom PMT array, and the minimum at large $r$, just below the gate grid. The results 
of the three measurements agree within 3\%.
The correction map obtained from the 40~keV line, which is used to correct the data, 
is shown in Fig.~\ref{fig::lce}.

\begin{figure}[tb]
  \includegraphics*[width=0.47\textwidth]{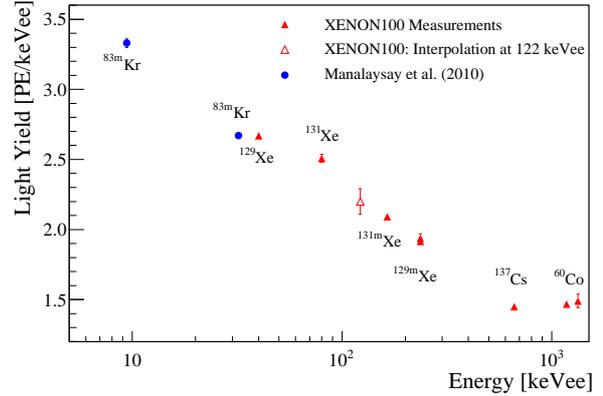}
  \caption{Measured light yield (in PE/keV$_\textnormal{\scriptsize ee}$) for different gamma lines 
used to calibrate
XENON100. A direct calibration at 122~keV$_\textnormal{\scriptsize ee}$ is not possible due 
to the limited penetration depth of $\gamma$-rays of this energy,
therefore the value has been interpolated using a fit to the available lines. 
The values at 9.4~keV$_\textnormal{\scriptsize ee}$ and 
32.1~keV$_\textnormal{\scriptsize ee}$ are taken
from \cite{ref::kr83_manalaysay} and scaled to the light yield at 122~keV$_\textnormal{\scriptsize ee}$ 
and the different electric field. They are given for comparison and not used in the fit.}\label{fig::ly}
\end{figure}

The light yield L$_y$ at 122~keV$_\textnormal{\scriptsize ee}$ is an important parameter to determine the nuclear recoil
equivalent energy scale from the S1 signal (see~\cite{ref::aprile2009,ref::manzur2010,ref::plante2011} 
and references therein). However, it cannot be measured directly in large detectors such as 
XENON100 since gamma rays of this energy do not penetrate into the fiducial volume. Instead, 
L$_y$ was determined from a fit to the measured volume-averaged scintillation yield 
for several calibration
peaks above and below 122~keV, namely 40~keV ($^{129}$Xe), 80~keV ($^{131}$Xe),
164~keV ($^{131m}$Xe), 
and 662~keV ($^{137}$Cs), where all peaks besides the last one are from inelastic
neutron interactions from the $^{241}$AmBe calibration. An empirical function, linear in $\log_{10}E$, was employed for the interpolation and results in a light yield 
of $\mathrm{L}_y=(2.20 \pm 0.09)$~PE/keV$_\textnormal{\scriptsize ee}$ at 122~keV
and the operating drift field of 0.53~kV/cm (see~Fig.~\ref{fig::ly}). Taking into account the measured
field quenching~\cite{ref::aprile2006,ref::doke2002,ref::kr83_manalaysay}, 
this corresponds to 4.3~PE/keV$_\textnormal{\scriptsize ee}$ at zero field at 122~keV.
The error of $\mathrm{L}_y$ includes the uncertainty due to the choice of fit function and takes into account potential variations in the volume-average, as the events from different sources used for the analysis are differently distributed throughout the TPC.

\subsection{Electric Field Correction}\label{sec::efieldCorr}

In order to reach the highest possible light collection efficiency and
therefore the lowest possible S1 energy threshold, the cathode mesh was optimized in terms of 
optical transparency using a mesh pitch of $\sim$5~mm. However, first measurements
of the $(r,z)$ distribution of events in the TPC revealed that the electric
field leaking through the cathode was underestimated in the design, leading
to an outward bending of the outermost field lines, close to the cathode.
There are no charge-insensitive regions from this effect, but it leads
to an inwards shift of the reconstructed event radius for 
$r>120$~mm and $z<-250$~mm. 

In order to optimize the detector response, a field correction was
determined by a numerical finite element calculation of the electric
field configuration, and cross checked with a simulation using the boundary 
elements method. The result
was verified with three independent measurements: 
the position of the outermost line of detected events, the
electron drift time distribution, as well as 
the uniform volume density of neutron activated xenon events 
at 164~keV and 236~keV from metastable states of $^{131}$Xe and $^{129}$Xe, respectively. 
A large amount of data with lines from activated
xenon was taken following a second $^{241}$AmBe calibration at the beginning of 2011, at a reduced extraction field to avoid PMT saturation effects. After applying the field correction, the homogeneity of
the event distribution can be confirmed within \mbox{2-3\%} up to $r=145$~mm, where the precision is limited by
the subtraction of the electromagnetic background.
This leads to a maximal uncertainty on the field correction which is equal to the position reconstruction uncertainty at the outermost fieldline. It decreases with smaller radii as the correction itself
becomes small. 

\begin{figure}[tb]
  \includegraphics*[width=0.47\textwidth]{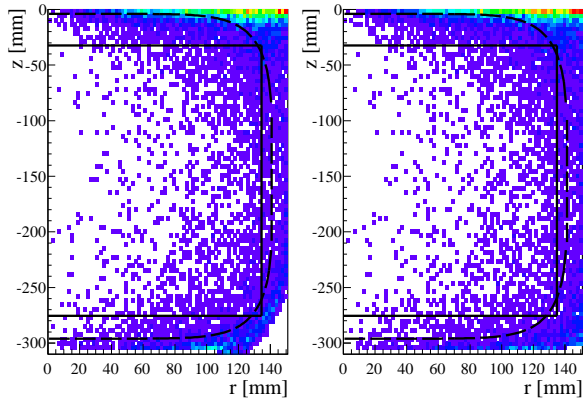}
  \caption{$(r,z)$ distribution of background events without (left) and with (right)
electric field correction. The non-uniform electric field
mostly affects the region at largest $r$ and $|z|$. The solid line indicates the 
volume of 40~kg fiducial mass used for the analysis presented in \cite{ref::xe100_11d}, the 
dashed line the 48~kg fiducial mass used in \cite{ref::xe100run08}.
}\label{fig::fieldcorr}
\end{figure}

All position-dependent S1 and S2 corrections can be obtained 
completely independent from the field correction. The field correction, however, improves the detector
performance for large fiducial masses and allows a more precise determination of 
the position of the event vertex for large $r$ and $z$. A comparison of this
detector region with and without field correction is shown in Fig.~\ref{fig::fieldcorr}
for the background data presented in Ref.~\cite{ref::xe100_11d}. 
As the correction is applied to both calibration and science
data, its uncertainty has negligible impact on dark matter results.

\subsection{Performance of the active LXe Veto}

The XENON100 TPC is completely surrounded by a volume of 99~kg LXe with a thickness
of $\sim$4~cm on the side of the PTFE cylinder, below the bottom PMTs, and above the diving bell. 
Due to the high stopping power of
LXe, it effectively suppresses background gamma rays from outer detector
materials and the passive shield. In order to improve the background
reduction even further, the LXe shield has been equipped with 64~PMTs.
32~PMTs monitor the side of the TPC, whereas two sets of 16~monitor the volume above and below.
The shield is optically separated from the target volume and is operated as an active veto.
This allows for the rejection of multi-scatter events when both LXe regions show S1
signals within $\pm$20~ns, where the time window is chosen to account for possible ADC
bit jitters and the slow component of LXe scintillation.

Due to the limited size of the detector vessel, the position of the PMTs in
the veto, and the boundary condition that all support structures and cables for
the TPC have to pass through the veto, there is a rather strong position
dependence of the veto response.  
Measurements with a collimated $^{137}$Cs source at more than
100~positions all around the detector were performed to determine the veto
detection thresholds as function of position. 
They were estimated from position and width of the measured 
$^{137}$Cs full absorption peak using a Monte Carlo simulation: The 90\%
detection thresholds are between 180~keV$_\textnormal{\footnotesize ee}$ and 235~keV$_\textnormal{\footnotesize ee}$ in the side veto,
130~keV$_\textnormal{\footnotesize ee}$ below the bottom PMT array, and between 10~keV$_\textnormal{\footnotesize ee}$ and 30~keV$_\textnormal{\footnotesize ee}$ above the
target volume. The threshold increases to about 450~keV$_\textnormal{\footnotesize ee}$ in regions behind veto PMTs. 

The energy deposition in the veto is slightly anti-correlated to the
energy deposition in the TPC and the active veto reduces the single-scatter gamma
background in the TPC very efficiently. From a detailed background study~\cite{ref::xe100_mc}
using these results, an
additional background reduction of 44\% and $>$70\% is obtained in the region of interest
for the full target and $\le$50~kg fiducial mass, respectively, compared to a
passive LXe shield.

\subsection{Combined Energy Scale and Energy Resolution}\label{sec::ces}

\begin{figure}[tb]
  \centering
  \includegraphics*[width=0.40\textwidth]{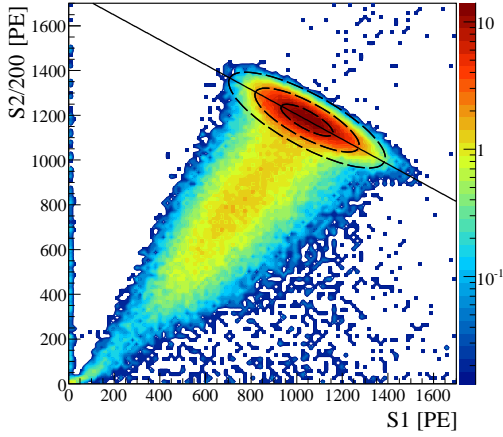}
  \caption{$^{137}$Cs calibration data in position-corrected S2-S1 parameter space. 
The signals are anti-correlated and a projection along the anti-correlation ellipse 
leads to an improved energy resolution. The color scale gives the number of events per bin.
}\label{fig::anticorr}
\end{figure}

Light and charge signal are anti-correlated
for interactions in LXe \cite{ref::ces_shutt,ref::ces_aprile,ref::ces_aprile2}. Every calibration line 
generates an ellipse in the S2-S1 plane, which can be described with a two-dimensional 
Gaussian in order to determine the anti-correlation angle~$\theta$, 
see~Fig.~\ref{fig::anticorr}. The projection of the peak along this angle allows for 
an improved energy resolution.

The angle~$\theta$ has been determined for the 40~keV$_\textnormal{\scriptsize ee}$ and 80~keV$_\textnormal{\scriptsize ee}$ peaks from inelastic neutron scattering, for 164~keV$_\textnormal{\scriptsize ee}$ and 236~keV$_\textnormal{\scriptsize ee}$ from metastable states of $^{131}$Xe and $^{129}$Xe, respectively, for the 662~keV$_\textnormal{\scriptsize ee}$ line from $^{137}$Cs, and for the $^{60}$Co lines at 1173~keV$_\textnormal{\scriptsize ee}$ and 1333~keV$_\textnormal{\scriptsize ee}$. 
It is quite constant for $E \gtrsim 100$~keV$_\textnormal{\scriptsize ee}$. The angles for the 40~keV$_\textnormal{\scriptsize ee}$ and 
80~keV$_\textnormal{\scriptsize ee}$ lines are smaller since the interaction is a combination of a nuclear recoil and a subsequent gamma emission. 

\begin{figure}[b!]
  \includegraphics*[width=0.47\textwidth]{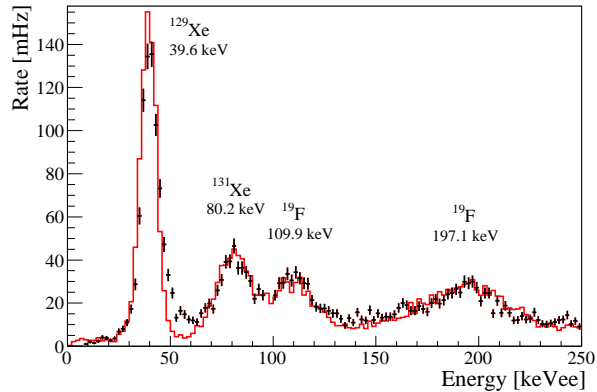}
  \caption{$^{241}$AmBe calibration spectrum in combined energy scale (electronic recoil interactions only). The spectrum agrees very well with the result from a Monte Carlo simulation (red line).
}\label{fig::ces}
\end{figure}
 
From the mean positions and angles obtained from calibration data, the combined energy scale for electronic recoil events has been defined. By comparing the resulting spectrum to Monte Carlo data, it was verified that the combined energy scale is indeed linear. Fig.~\ref{fig::ces} compares the measured electronic recoil spectrum from the calibration with $^{241}$AmBe, using the combined energy scale, with a Monte Carlo generated spectrum. This scale is currently only used for background studies \cite{ref::xe100_mc}; the WIMP search data is analyzed using a S1-based nuclear recoil energy scale \cite{ref::xe100_11d,ref::xe100run08}.

\begin{figure}[tb]
  \includegraphics*[width=0.47\textwidth]{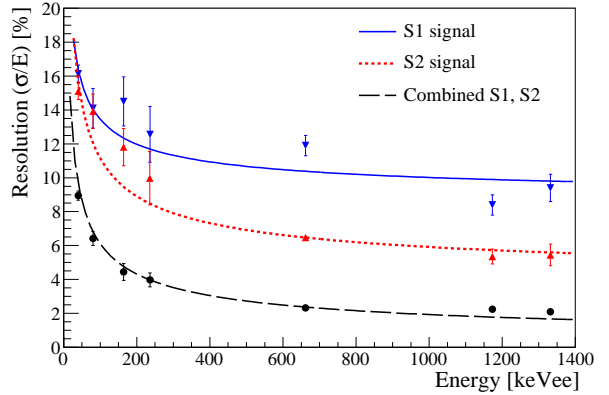}
  \caption{Measured resolution ($\sigma/E$) of gamma calibration lines between 40~keV$_\textnormal{\scriptsize ee}$ and 1333~keV$_\textnormal{\scriptsize ee}$ in S1, S2, and combined energy scale, together with fits describing the $1/\sqrt{E}$ dependence.}\label{fig::eres_lines}
\end{figure}

\begin{figure}[b!]
  \includegraphics*[width=0.47\textwidth]{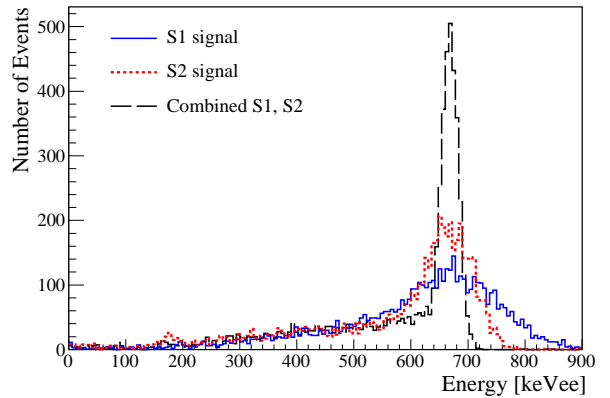}
  \caption{Spectrum of $^{137}$Cs at 662~keV$_\textnormal{\scriptsize ee}$ in three different energy scales: The resolution (1~$\sigma$) is 12.5\% for the S1 scale, 6.5\% for the S2 scale, and 2.3\% for the combined energy scale where the anti-correlation between charge and light signal is exploited. Only single-scatter events are considered in the plot and a veto cut is applied, effectively reducing the Compton continuum.
}\label{fig::eres}
\end{figure}

The energy resolution has been determined from the same gamma calibration lines for three different energy scales, based on the S1 signal alone, the S2 signal alone, and the combined energy scale, see Fig.~\ref{fig::eres_lines}. All position-dependent corrections have been applied. The energy dependence of the resolution $\sigma/E$ can in all cases be described by
\begin{equation}
\frac{\sigma(E)}{E} = \frac{c_1}{\sqrt{E}}+c_2 \textnormal{,} 
\end{equation}
where $E$ is the gamma energy and the $c_i$ are constants that are different for the three scales. $c_2$ is compatible with zero for the combined energy scale. At 1~MeV, the resolution is 12.2\% for the S1-based scale, 5.9\% for the S2-based scale, and 1.9\% in the combined energy scale. The S2-based value is comparable to the resolution of NaI(Tl) crystals and the resolution of the combined
energy scale is even better, in particular when compared to large crystals. Fig.~\ref{fig::eres} shows the change of the $^{137}$Cs spectrum using the three scales. For this figure, a veto cut is applied and only events with one interaction in the TPC are considered in order to suppress the Compton continuum.